%% file: main.tex
\documentclass[aps,prl,reprint,showpacs,superscriptaddress]{revtex4-2}
\usepackage{graphicx}
\usepackage{amsmath}
\usepackage{amssymb}
\usepackage{amsfonts}
\usepackage{dcolumn}
\usepackage{xcolor}
\usepackage{subfigure}
\usepackage{float}
\usepackage{bm}
\usepackage{amsthm}
\usepackage{gensymb}
\usepackage[T1]{fontenc}
\usepackage[mathscr]{euscript}
\usepackage{bbm}
\DeclareMathAlphabet{\mathcalligra}{T1}{calligra}{m}{n}
\usepackage{dcolumn}
\usepackage{capt-of}
\newcommand{\chb}{\textcolor{blue}}

\usepackage{pifont}% http://ctan.org/pkg/pifont
\newcommand{\cmark}{\ding{51}}%
\newcommand{\xmark}{\ding{54}}%

\usepackage{color} % use colored text in latex

\usepackage{hyperref}
\hypersetup{
colorlinks=true,final=true,
        linkcolor=blue,
        citecolor=blue,
        filecolor=blue,
        urlcolor=magenta,
}
\usepackage{bigints}   
\usepackage{epstopdf}   
\usepackage{soul} % for todo and strikethrough

\begin{document}

\title{Prediction of Room Temperature Electric Field Reversal of Magnetization in the Family of {$A_4B_3\rm{O}_9$} Layered Oxides}
\author{Urmimala Dey}
\affiliation{Centre for Materials Physics, Durham University, South Road, Durham DH1 3LE, United Kingdom}
\affiliation{Luxembourg Institute of Science and Technology (LIST), Avenue des Hauts-Fourneaux 5, L4362, Esch-sur-Alzette, Luxembourg}
\author{Emma E. McCabe}
\affiliation{Centre for Materials Physics, Durham University, South Road, Durham DH1 3LE, United Kingdom}
\author{Jorge \'{I}\~{n}iguez-Gonz\'{a}lez}
\affiliation{Luxembourg Institute of Science and Technology (LIST), Avenue des Hauts-Fourneaux 5, L4362, Esch-sur-Alzette, Luxembourg}
\affiliation{Department of Physics and Materials Science, University of Luxembourg, 41 Rue du Brill, L4422, Belvaux, Luxembourg}
\author{Nicholas C. Bristowe}
\email{nicholas.bristowe@durham.ac.uk}
\affiliation{Centre for Materials Physics, Durham University, South Road, Durham DH1 3LE, United Kingdom}

\date{\today}

\begin{abstract}
The promise of a strong magnetoelectric coupling in a multiferroic material is not only of fundamental interest, but also forms the basis of next generation memory devices where the direction of magnetization can be reversed by an external electric field. 
Using group-theory led first-principles calculations, we have identified a hitherto unknown polar phase of the {$A_4B_3\rm{O}_9$} layered oxides, where the polar mode couples to the magnetic modes through a rare $\Gamma$-point magnetoelectric-multiferroic coupling scheme such that the net magnetization can be directly reversed by an electric field switching of the polar mode. Furthermore, in agreement with previous experimental observations, we predict room temperature magnetism in {$A_4B_3\rm{O}_9$} oxides which indicates the promising practical applications of these compounds in the next generation memory devices.
\end{abstract}

\maketitle

\chb{\textit{Introduction. --- }} 
Magnetoelectric-multiferroic (ME-MF) materials with intrinsic cross-coupling between electrical and magnetic order parameters are promising for the next generation memory devices where an external electric field can switch the direction of magnetization leading to enhanced speed and reduced power consumption~\cite{Fiebig2004,Spaldin2006,Ederer2006,Fiebig2016,Spaldin2019,Fert2024}. So far, despite intensive research efforts, only a handful of bulk materials with electric field switchable magnetization have been predicted and observed in experiments~\cite{Fennie2008,Fennie2011,RFO2012,Chai2014,Fert2024,Kocsis2019}, and unfortunately none that order at room temperature (RT). Therefore, the search for ME-MF materials with RT electric field control of magnetization remains of utmost importance for realistic applications in memory devices~\cite{Ramesh2021,Fert2024}.  \begin{figure}[h!]
\includegraphics[scale=0.375]{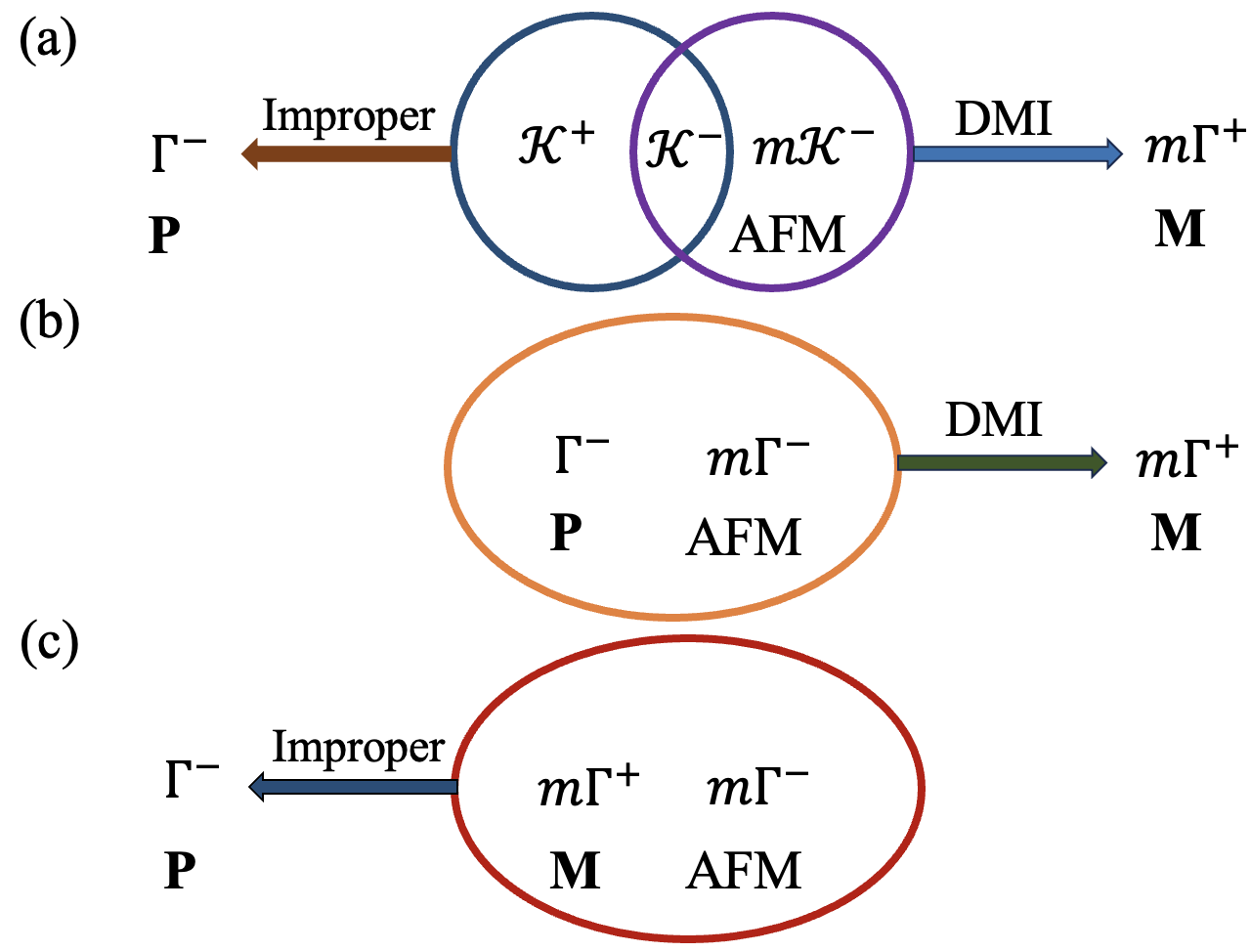}
 \caption{Trilinear coupling terms in the free energy expansion representing (a) $\mathscr{K}$-point ME-MF scheme in perovskites involving polarization and magnetization~\cite{Senn2018}, where $\mathscr{K}$ denotes a general zone-boundary point. $m\mathscr{K}^-$ is the irrep associated with AFM ordering, whereas $\mathscr{K}^-$ and $\mathscr{K}^+$ are the irreps representing nonpolar and nonmagnetic order parameters. (b) $\Gamma$-point coupling scheme allowing for the reversal of $\textbf{M}$ with $\textbf{P}$ in systems with larger unit cells where AFM order parameters transform as $m\Gamma^-$ irreps. (c) The same $\Gamma$-point ME-MF scheme in type-II MFs where the AFM mode couples with $\textbf{M}$ to induce improper ferroelectricity.}
\label{Gamma-pt}
\end{figure} 

Such control of magnetization can be achieved in a (single phase) antiferromagnetic (AFM) ME-MF material where weak ferromagnetism (wFM) arises due to canting of the collinear AFM spins via the Dzyaloshinskii-Moriya interaction (DMI)~\cite{DMI1958,Moriya1960,Scott2011}. Depending on the crystal structure and underlying symmetries, electric polarization ($\textbf{P}$) can couple to the net magnetization ($\textbf{M}$) in these materials via different mechanisms~\cite{Fennie2008,Fennie2011,Yang2014}. The advantage of these type of strategies is that, since the couplings are at odd order (sometimes called `improper' couplings), switching $\textbf{P}$ can necessarily switch $\textbf{M}$. Inspired by these ME schemes, Senn and Bristowe have enumerated the possible ME-MF couplings in perovskites using a group-theoretical approach~\cite{Senn2018}. However, since AFM orderings in perovskite systems are always described by zone-boundary irreducible representations (irreps) of the parent space group, even the lowest-order ME-MF coupling schemes must involve a two-step process, each step contributing odd order energy terms in $\textbf{P}$ and $\textbf{M}$ separately with codependent order parameters~\cite{Senn2018}, as shown in Fig.~\ref{Gamma-pt}(a).   

More generally, however, it should be possible to construct a simpler $\Gamma$-point scheme where $\textbf{P}$ and $\textbf{M}$ couple with a $\Gamma$-point AFM ordering mode in an `improper' manner such that the crystal momenta, inversion symmetry and the time reversal symmetry are preserved (see Figs.~\ref{Gamma-pt}(b)$-$(c)). When $\textbf{P}$ and AFM ordering modes couple to give rise to a net $\textbf{M}$ induced by the DMI, application of an electric field can directly reverse $\textbf{M}$ via the reversal of $\textbf{P}$, since the primary AFM order parameter is less likely to switch due to magnetic anisotropy~\cite{Fennie2008,Spaldin2006,Ederer2005}. This simpler $\Gamma$-point ME-MF scheme, shown in Fig.~\ref{Gamma-pt}(b), allows for the switching mechanism to be contained in only one trilinear term but likely requires proper ferroelectricity and has rarely been observed~\cite{Fennie2008,Ederer2008} to the best of our knowledge. On the other hand, the same $\Gamma$-point coupling in Fig.~\ref{Gamma-pt}(c) can also explain the induction of improper ferroelectricity observed in some type-II MFs~\cite{BMO2022,Dey2023} where the AFM mode couples with $\textbf{M}$ to break the spatial inversion symmetry of the system, though this usually requires complex magnetic structures. 

Using first-principles density functional theory (DFT) calculations guided by group-theoretical analysis, we identify a hitherto unknown polar phase of bulk {$A_4B_3\rm{O}_9$} layered oxides ($A$: rare-earth and/or alkali-earth cations; $B$: Co, Ni, Fe) where an applied electric field can switch the magnetization between 180\degree~ symmetry equivalent states through the $\Gamma$-point ME-MF scheme shown in Fig.~\ref{Gamma-pt}(b). Full computational details are given in Supplemental Material (SM)~\cite{supp}.   
Previous experimental studies on these layered oxides demonstrated long-range AFM ordering of the spins above RT~\cite{Rahmani2022,LaCo1998}, indicating the possibility of RT electric field switching of magnetization in {$A_4B_3\rm{O}_9$} layered oxides, which is further supported by our calculations of magnetic exchange interaction parameters. We explain the design principles to stabilize the polar phase with nontrivial ME effect starting from an otherwise nonpolar structure without net magnetization, in the hope of inspiring future experimental work.  

\begin{figure}[ht!]
\includegraphics[scale=0.36]{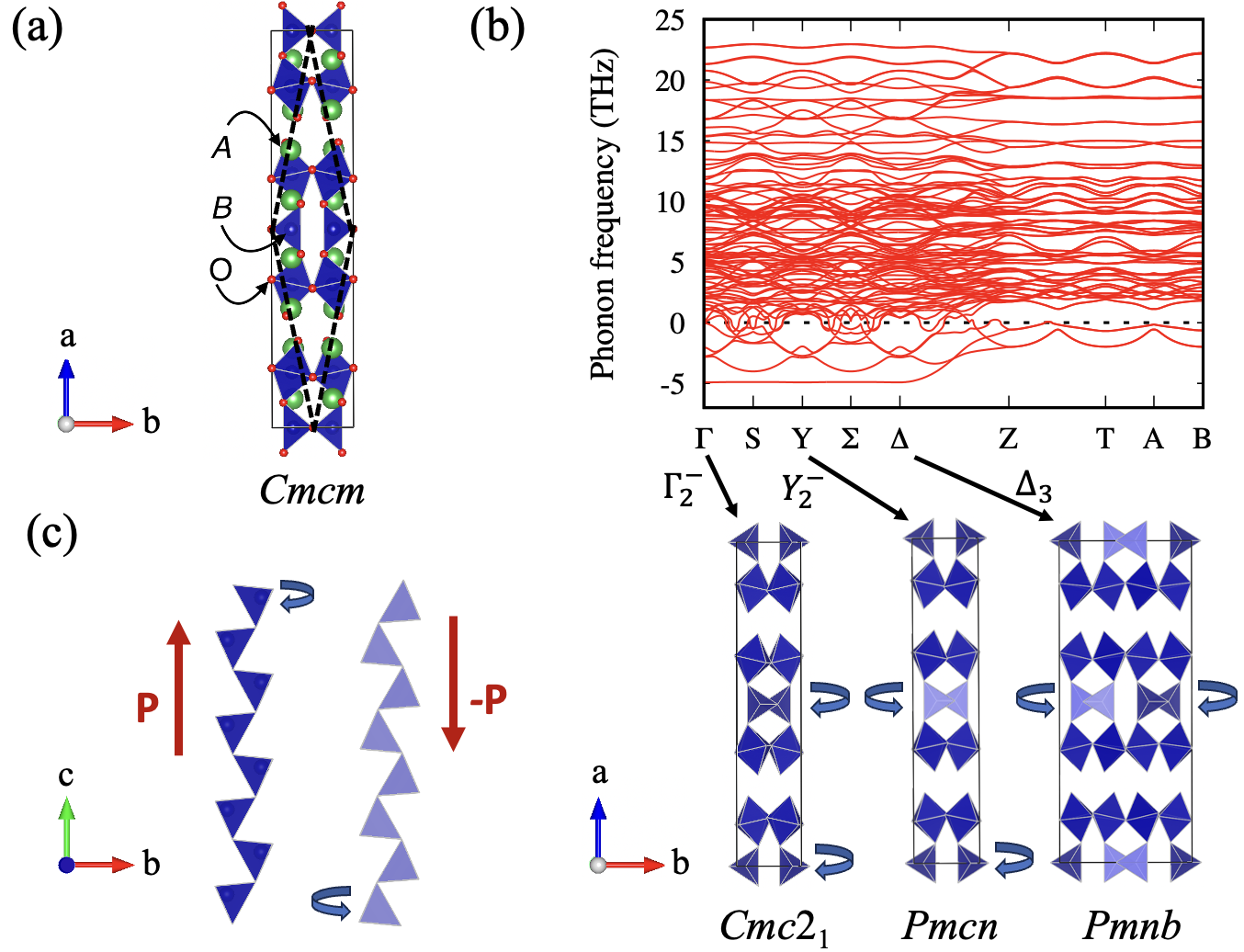}
    \caption{(a) Crystal structure of the high-symmetry $Cmcm$ phase of the {$A_4B_3\rm{O}_9$} layered oxides consisting of planes of $B$O$_6$ octahedra and $B$O$_4$ tetrahedra alternating along the out-of-plane direction. Conventional and primitive unit cells are shown by solid and dashed lines, respectively. (b) Phonon spectrum calculated (in the primitive basis) for the $Cmcm$ structure of Nd$_4$Co$_3$O$_9$ as a representative of the {$A_4B_3\rm{O}_9$} layered oxides. The phonon dispersion shows a number of instabilities at the zone center and zone boundary points, which when condensed, result in a variety of structural variants arising from the cooperative rotations of the tetrahedral units. (c) In-plane view of the right- and left-handed tetrahedral chains with opposite polarizations.}
\label{fig:structures}
\end{figure}

\chb{\textit{Results. --- }} 
Fig.~\ref{fig:structures} shows the crystal structure of the {$A_4B_3\rm{O}_9$} layered oxides that contains slabs of (oxygen-deficient) perovskite-type $AB$O$_3$ triple layers translated relative to each other in the basal plane, and separated by rocksalt-type $A$O layers~\cite{LaCo1998,NdSr2005}, similar to the $n = 3$ Ruddlesden-Popper phases~\cite{RP3}. $A$-sites are occupied by rare-earth and/or alkali-earth elements while the $B$-sites are occupied by transition metal ions. Oxygen vacancies are formed at the central perovskite layer in each block in an ordered manner giving rise to planes of $B$O$_6$ octahedra and $B$O$_4$ tetrahedra alternating along the out-of-plane direction reminiscent of those in the brownmillerite (BM) phases~\cite{BM}. 

{$A_4B_3\rm{O}_9$} layered oxides appear relatively understudied in literature compared to other oxygen-deficient layered compounds like  BMs~\cite{BM2008,Rondinelli2015,Bellaiche2019,Bellaiche2023} and Grenier phases~\cite{Grenier1976,Luo2013,Shin2023,Zhang2024}.
The parent phase of these layered oxides has $Cmcm$ symmetry (no. 63), which contains no rotations of the tetrahedra but allows for tilting of the $B$O$_6$ octahedra, as shown in Fig.~\ref{fig:structures}(a). The parent $Cmcm$ structure can also be described as a disordered phase with no long-range ordering of the tetrahedra. Previous experiments on $Ln$Sr$_3$Fe$_3$O$_9$ ($Ln$: La, Pr, Nd) show that these layered oxides crystallize into an average structure of $Cmcm$ symmetry without any signature of long-range ordering of the tetrahedral chains~\cite{LaSr2019,LaSr2016,NdSr2005,PrSr2012}. However, the tetrahedra in these layered compounds can order by rotating about the out-of-plane axis in clockwise and anticlockwise directions leading to the formation of right- and left-handed apex-linked 1D tetrahedral chains with distinct chiralities, analogous to the BMs and Grenier structures. Since the tetrahedra are corner-connected, rotation about the out-of-plane direction is cooperative i.e. rotation of one tetrahedron (e.g., clockwise) about the out-of-plane direction causes the nearest corner-connected in-plane tetrahedra to rotate in the opposite direction (e.g., anticlockwise). 

Depending on the relative ordering of the tetrahedral chains within the unit cell, it is possible to observe a number of structurally diverse phases~\cite{supp}. When all the tetrahedra rotate in the same direction (either left-handed or right-handed), the dipole moments arising from the displacement of cations away from the center of each tetrahedron add up, resulting in a polar structure in the $Cmc2_1$ space group (no. 36) that is associated with a polar mode transforming as the $\Gamma^-_2(a)$ irrep of the parent $Cmcm$ phase (see Fig.~\ref{fig:structures}(b)). Note that in the $Cmc2_1$ structure, electric polarization is in-plane and along the tetrahedral chain direction. 
On the other hand, if the tetrahedra rotate in opposite senses in successive layers (i.e. intralayer polarization reversing from layer to layer), the dipole moments from each layer cancel out forming an antipolar structure as depicted in Fig.~\ref{fig:structures}(b). This antipolar structure is related to the $Cmcm$ phase by the $Y^-_2(a)$ irrep that reduces the symmetry from $Cmcm$ to $Pmcn$ (no. 62, $bca$ setting).  
A neutron diffraction study by Hansteen \textit{et al.} on La$_4$Co$_3$O$_9$ reveals the formation of long range ordering of the CoO$_4$ tetrahedra along the out-of-plane direction leading to an antipolar phase in the $Pnma$ space group (no. 62)~\cite{LaCo1998}. 

It is interesting to note that different intralayer and interlayer tetrahedral twisting patterns can further lead to a variety of other distinct phases resulting in superstructures with longer periods, as observed in a local scale in Ca$_4$Fe$_2$Mn$_{0.5}$Ti$_{0.5}$O$_9$~\cite{CaFe2011}. 
An example of a superstructure with intralayer switching of tetrahedral rotation patterns described by $Pmnb$ symmetry (no. 62) is shown in Fig.~\ref{fig:structures}(b), which is associated with the $\Delta_3(a,-a)$ irrep of the parent phase. 
We have identified another novel polar phase with $Pmc2_1$ symmetry which contributes an energy
term of the form $Q^2_{\Delta_3}Q_{\Gamma^-_2}Q_{Y^-_2}$ in the free energy expansion of the high symmetry phase (Section S2 F in ~\cite{supp}). This kind of novel ferroelectric phase with a quadratic-bilinear coupling has also been observed in BM oxides~\cite{Tian2018}.

Indeed, our phonon calculations for the paraelectric $Cmcm$ structure of $Ln_{4}$$B$$_3$O$_{9}$ ($Ln$: La, Pr, Nd; $B$: Co, Ni) and La$A^\prime_3$Fe$_3$O$_{9}$ ($A^\prime$: Sr, Ca) compounds reveal a number of instabilities at the zone center and zone boundary points, which when condensed, result in a variety of structural variants arising from the cooperative rotations of the tetrahedral units, as shown in Fig.~\ref{fig:structures}(b). 
The phonon spectra calculated for the $Cmcm$ phase of all the considered $A_{4}B_3$O$_{9}$ systems show qualitatively similar features with a strongly unstable flat phonon branch along the $\Gamma-$S$-\Sigma-$Y$-\Delta$ direction in the Brillouin zone related to the tetrahedral chain ordering distortions. Presence of a flat phonon band indicates that the different structural variants derived from these instabilities will be close in energy (at the harmonic level, at least).

Starting with the optimized structures, we compute the energy of the nonpolar $Cmcm$ phase relative to the polar $Cmc2_1$ phase ($\Delta E_{\rm{NP}}$) across three sets of compounds containing different $B$-site cations ($B$ = Co, Ni and Fe). We find that all the considered materials are insulating with a band gap in the range of 1.2 to 2.4 eV~\cite{supp}. In each series, inclusion of smaller $A$-site cations is found to increase the degree of tetrahedral rotations leading to higher stability of the low-symmetry phases as shown in Table~\ref{Table1}. 
\begin{table}[h]
  \centering
  \caption{Energetics and band gaps of the considered {$A_4B_3\rm{O}_9$} layered oxides. Here, $\Delta E_{\rm{NP}}$ is the energy of the nonpolar $Cmcm$ phase relative to the polar $Cmc2_1$ phase. $\Delta E_{\rm{PA}}$  and $Q_{\Gamma^-_2}$ denote the relative energy of the polar $Cmc2_1$ phase with respect to the antipolar $Pmcn$ phase and the amplitude of the polar mode, respectively. Band gaps are determined for the lowest energy phases, while $Q_{\Gamma^-_2}$ is calculated for the relaxed $Cmc2_1$ structure of each compound.} 
  \label{Table1}
  \begin{tabular}{|c|c|c|c|c|c|}
    \hline
    \hline\rule{0pt}{1.0\normalbaselineskip}
   $B$-site &Layered& $\Delta E_{\rm{NP}}$ &$\Delta E_{\rm{PA}}$&$Q_{\Gamma^-_2}$ &Band gap\\
   element&oxide&(eV/f.u.)&(meV/f.u.)&({\AA})&(eV) \\
    \hline
     \rule{0pt}{1.0\normalbaselineskip}
      &La$_4$Co$_3$O$_9$&0.63&0.42&1.67&2.0\\
       \rule{0pt}{1.0\normalbaselineskip}
     Co&Pr$_4$Co$_3$O$_9$&1.09&0.11&1.72&1.9\\
      \rule{0pt}{1.0\normalbaselineskip}
     &Nd$_4$Co$_3$O$_9$&1.27&$-3.21$&2.22&2.4 \\
     \hline\rule{0pt}{1.2\normalbaselineskip}
     &La$_4$Ni$_3$O$_9$&0.55&$-0.15$&1.60&1.3\\
     \rule{0pt}{1.0\normalbaselineskip}
    Ni&Pr$_4$Ni$_3$O$_9$&0.73&$-0.32$&1.65&1.2\\
     \rule{0pt}{1.0\normalbaselineskip}
     &Nd$_4$Ni$_3$O$_9$&0.76&$-0.71$&1.68&1.2 \\
     \hline\rule{0pt}{1.2\normalbaselineskip}
     Fe&LaSr$_3$Fe$_3$O$_9$&0.44&0.003&1.70&1.9\\
      \rule{0pt}{1.0\normalbaselineskip}
      &LaCa$_3$Fe$_3$O$_9$&0.74&$-9.10$&1.75&1.7\\
    \hline
    \hline
\end{tabular}
\end{table}

In Table~\ref{Table1}, we tabulate the energy of the polar $Cmc2_1$ phase relative to the antipolar $Pmcn$ phase  ($\Delta E_{\rm{PA}}$) for the considered layered compounds. The antipolar phase is the ground state (g.s.) for La$_4$Co$_3$O$_9$ in agreement with earlier experimental results~\cite{LaCo1998}. However, as we decrease the $A$-cation size from La$^{3+}$ to Nd$^{3+}$, the antipolar phase becomes metastable and the g.s. acquires finite polarization with $Cmc2_1$ symmetry. A similar trend in $\Delta E_{\rm{PA}}$ is observed in the Ni- and Fe-series.
Furthermore, the amplitude of the polar mode (see Section S2 D in~\cite{supp} for the definition) computed for the fully relaxed $Cmc2_1$ structure of each compound shows that inclusion of smaller $A$-site cations increases the polar distortion.
Note that this trend in $\Delta E_{\rm{PA}}$ is opposite in BMs where larger tetrahedral chain rotations favor the antipolar phase~\cite{Parsons2009}. In the case of LaSr$_3$Fe$_3$O$_9$, $\Delta E_{\rm{NP}}$ is smaller and the polar and antipolar phases have almost identical energies ($\Delta E_{\rm{PA}} \sim 0.003$ meV/f.u.), implying that there might not be any long range ordering of the tetrahedral chains~\footnote{Results in Table~\ref{Table1} are given only for the electron doped models of La$A^\prime_3$Fe$_3$O$_9$ ($A^\prime$ = Ca, Sr) and as shown in SM~\cite{supp}, relative energies of the polar and antipolar phases of La$A^\prime_3$Fe$_3$O$_9$ vary depending on the cation ordering model considered.}, consistent with the previous experimental observation of an average $Cmcm$ structure~\cite{LaSr2019,LaSr2016}

We next check the stability of the polar $Cmc2_1$ structure of the compounds with negative $\Delta E_{\rm{PA}}$ against the formation of superstructures with longer periods of tetrahedral twisting patterns~\cite{supp}. 
Focusing on the Co-series (due to the challenges of modeling $A$-cation disorder with DFT in the Fe-series, and with the Ni-series not yet synthesized), we find that polar $Cmc2_1$ phase is the g.s. of Nd$_4$Co$_3$O$_9$. Dynamical stability of the $Cmc2_1$ structure of Nd$_4$Co$_3$O$_9$ is shown in Section S2 F of SM~\cite{supp}.

{$A_4B_3\rm{O}_9$} compounds are shown to possess above RT magnetic order on the transition metal lattice in earlier experiments~\cite{LaSr2019,NdSr2005,Rahmani2022,LaCo1998}. Previous neutron diffraction studies on $Ln$Sr$_3$Fe$_3$O$_9$ ($Ln$: La, Pr, Nd)~\cite{LaSr2019,NdSr2005} and La$_4$Co$_3$O$_9$~\cite{LaCo1998} have also characterized the g.s. magnetic structures of these materials where the spins are found to order antiferromagnetically above RT along the tetrahedral chain direction. Considering four collinear spin configurations of the $B$-site cations, %in the lowest energy phases given in Table~\ref{Table1}, 
namely, ferromagnetic and A-, C-, and G-type AFM spin arrangements, we first confirm the G-AFM structure
as the g.s. for the Co-series compounds as observed in earlier experiments~\cite{LaCo1998}. 
Noncollinear magnetic calculations including spin-orbit coupling further reveal that all the systems have magnetic easy axis along the tetrahedral chain direction i.e. in the plane along the [001] direction in agreement with previous experimental observations~\cite{LaSr2019,LaCo1998}. 
Details of the g.s. crystal structures and magnetic configurations are given in SM~\cite{supp}.
\begin{figure}[b]
\includegraphics[scale=0.262]{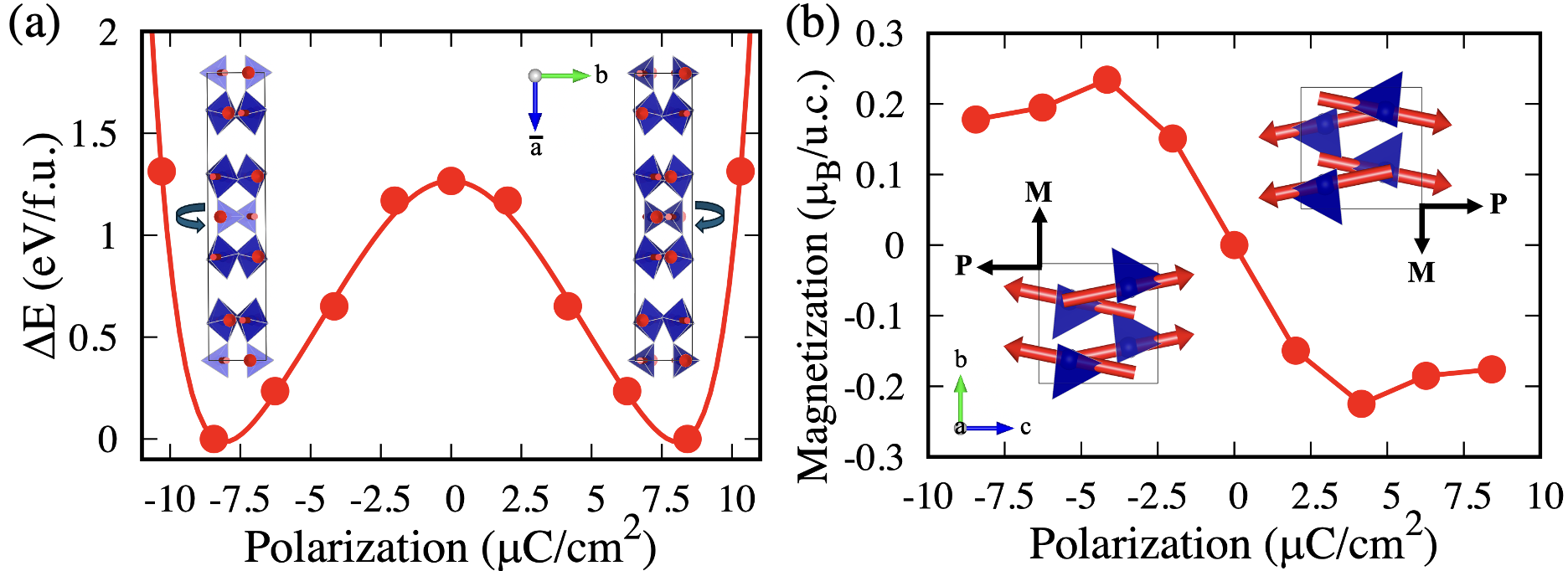}
\caption{(a) Polarization switching path in Nd$_4$Co$_3$O$_9$ calculated for the polar $Cmc2_1$ phase with G-AFM magnetic configuration. As seen, the sense of rotations of the tetrahedra is flipped as we go from the $+\textbf{P}$ states to the $-\textbf{P}$ states. (b) Calculated net magnetization as a function of the polar mode amplitude demonstrating the reversal of magnetization with the reversal of the electric polarization. Spin moments are denoted by red arrows. Note that only the magnetic moments on the tetrahedral sites are shown for clarity.} 
\label{switching_path}
%\twocolumngrid
\end{figure}

Interestingly, we find that in the $Cmc2_1$ phase, the polar mode induces a net magnetization perpendicular to the tetrahedral chain direction (i.e. along [010]) via a canting of the antiferromagnetically ordered (C- and G-AFM) collinear spins of the $B$-site cations along the chain direction (i.e. along [001]), which is otherwise absent in the paraelectric $Cmcm$ phase. Both C- and G-AFM spin configurations in the $Cmc2_1$ structure correspond to the magnetic space group $Cm^\prime c2^\prime_1$ (magnetic point group $m^\prime m2^\prime$) which is related to the high symmetry $Cmcm$ space group by a polar distortion transforming as the $\Gamma^-_2(a)$ irrep and two magnetic distortions transforming as $m\Gamma^-_3(a)$ and $m\Gamma^+_4(a)$ irreps. The $m\Gamma^-_3$ irrep is associated with the primary AFM magnetic ordering, whereas the $m\Gamma^+_4$ irrep represents the secondary wFM mode. Group theoretical analysis reveals that these three modes couple in an `improper' manner, contributing a trilinear energy term of the form: 
\begin{equation*}
    \mathcal{F}=\gamma Q_{\Gamma^-_2} Q_{m\Gamma^+_4} Q_{m\Gamma^-_3}
\end{equation*} 
in the free energy expansion of the parent phase, where $\gamma$ is the expansion coefficient.

As a consequence of this $\Gamma$-point coupling, reversal of \textbf{P} via an electric field would necessarily require reversal of just one of the two magnetic modes. We have investigated this point explicitly via first-principles simulations showing that, upon reversal of the polar distortion, the spins immediately evolve toward a configuration in which the wFM mode switches while the G-AFM order remains the same (see Section S2 I in Ref. \cite{supp}). Hence we expect $180\degree$ reversal of \textbf{M} with an applied electric field. 

Fig.~\ref{switching_path}(a) shows the polarization double well in Nd$_4$Co$_3$O$_9$ calculated for the g.s. G-AFM magnetic configuration in the polar $Cmc2_1$ phase with polarization along [001]. Here, we plot the relative energy of the $Cmc2_1$ structure ($\Delta E$) as a function of the polar mode amplitude ($Q_{\Gamma^-_2}$). 
As seen, $\Delta E$ is maximum for the $Cmcm$ phase that corresponds to \textit{P} = \textit{M} = 0. As we increase $Q_{\Gamma^-_2}$, $\Delta E$ decreases on both sides reaching a minimum at the relaxed value of $Q_{\Gamma^-_2}$ and corresponds to $P^0$ $\sim$$\pm 8.41$ $\mu$C/cm$^2$ which is about $2-4$ times larger than the electric polarization in BMs~\cite{Bellaiche2019,Bellaiche2023,Tian2018}. Fig.~\ref{switching_path}(b) shows the resulting net magnetization for the different amplitudes of the polar mode. For small $P$, the magnetization varies linearly with $Q_{\Gamma^-_2}$, which is consistent with the $\Gamma$-point scheme shown in Fig.~\ref{Gamma-pt}(b) and reaches a value of $M^0$ $\sim$$\pm 0.18$ $\mu_{\rm B}$ per unit cell for the relaxed value of $Q_{\Gamma^-_2}$. Interestingly, the spin canting angles are flipped as we reverse the electric polarization resulting in opposite net magnetization $-\textbf{M}$ and $+\textbf{M}$ in the $+\textbf{P}$ and $-\textbf{P}$ states, respectively, demonstrating the switching of magnetization via the reversal of the polar mode. 
Our first-principles calculations thus predict Nd$_4$Co$_3$O$_9$ as an ideal candidate for observing the electric field switching of magnetization. 

Finally, to investigate the possibility of RT magnetization switching in Nd$_4$Co$_3$O$_9$, we calculate the magnetic exchange parameters ($J$'s) of Nd$_4$Co$_3$O$_9$ and compare them with the $J$'s of La$_4$Co$_3$O$_9$ where RT long range magnetic order has been observed experimentally~\cite{LaCo1998,Rahmani2022}. We find that the calculated $J$'s for the $Pmcn$ structure of La$_4$Co$_3$O$_9$ are almost identical to those computed for the g.s. $Cmc2_1$ structure of Nd$_4$Co$_3$O$_9$ (Section S2 J in~\cite{supp}), indicating that Nd$_4$Co$_3$O$_9$ is also likely to exhibit long range magnetic ordering at RT. 

\chb{\textit{Discussion and conclusions. --- }}
We have proposed a novel $\Gamma$-point ME-MF scheme that allows for the reversal of DMI-induced magnetization with an external electric field and has rarely been observed~\cite{Fennie2008,Ederer2008} to the best of our knowledge. 

Our first-principles calculations reveal a variety of previously unknown structurally distinct phases of the {$A_4B_3\rm{O}_9$} oxides arising from the cooperative rotations of the tetrahedral units. In particular, a polar structure with $Cmc2_1$ symmetry has been identified that allows for a 180\degree ~reversal of magnetization by an electric field switching of the polar mode through the $\Gamma$-point ME-MF scheme. We find that smaller $A$-site cations can induce a larger polar distortion stabilizing the polar $Cmc2_1$ structure with net magnetization. 

The calculated values of $\Delta E_{\rm{NP}}$ which generally measures the upper bound of the switching barrier~\cite{Bellaiche2023} are in the range of 0.44 eV to 1.27 eV per formula unit and of the same order of magnitude calculated for similar geometric ferroelectrics~\cite{Mulder2013,Kang2019,Bellaiche2023}.  
Previous experimental evidence of ferroelectric switching in thin films of BMs with similar magnitudes of $\Delta E_{\rm{NP}}$~\cite{Kang2019} suggests the possibility of experimental observation of electric field reversal of polarization in {$A_4B_3\rm{O}_9$} layered oxides. However, one should note that while these layered oxides draw similarities with BMs, the BMs have not been found to allow for a 180\degree ~ME-MF switching scheme.

From our first-principles calculations, we put forward Nd$_4$Co$_3$O$_9$ with G-AFM ordered $Cmc2_1$ g.s. as an ideal candidate to observe the ME-MF switching. Moreover, in analogy with previous experimental observations, we predict RT electric field reversal of magnetization in Nd$_4$Co$_3$O$_9$ which indicates the possibility of its practical applications in the next generation memory devices. Note that magnetic anisotropy is crucial to a robust switching mechanism and must be optimized in future material design. Because of the similar ionic radii of La$^{3+}$ (1.36 \AA, C.N. 12) and Nd$^{3+}$ (1.27 \AA, C.N. 12), the Nd analog of LaSr$_3$Fe$_3$O$_9$ has been successfully synthesized in previous experiments~\cite{NdSr2005}. Therefore, it is likely that the Nd-substituted analog of La$_4$Co$_3$O$_9$ can also be synthesized. Moreover, Olafsen \textit{et al.} have demonstrated the preparation of the oxidized compound Nd$_4$Co$_3$O$_{10}$~\cite{NdCoO10}, showing the possibility of synthesizing its reduced derivative.

In summary, we have identified a series of new layered oxides that have all the necessary ingredients for the elusive control of magnetization with an electric field at RT. Apart from important memory applications, these materials show a range of intriguingly subtle magnetic and structural phases, apparently coupled to their layered and vacancy ordered nature. This understanding will likely provide new opportunities within related active fields on Ruddlesden-Popper and BM materials such as memristors~\cite{Lu2017,Mou2021}, catalysis~\cite{Forslund2018,Liu2020}, photoferroics~\cite{Blancon2020,Tsai2016}, and 2D magnetism~\cite{Dashwood2023,Markovic2020}.

\chb{\textit{Acknowledgments. --- }}
U.D. and N.C.B. acknowledge the Leverhulme Trust for a research project grant (Grant No. RPG-2020-206).
This work made use of the facilities of the N8 Centre of Excellence in Computationally Intensive Research (N8 CIR) provided and funded by the N8 research partnership and EPSRC (Grant No. EP/T022167/1). The Centre is coordinated by the Universities of Durham, Manchester and York. This work also used the ARCHER2 UK National Supercomputing Service (https://www.archer2.ac.uk)~\cite{archer2} and the Hamilton HPC Service of Durham University. We acknowledge useful discussions with Mark S. Senn. 

\chb{\textit{Data availability. --- }}
The authors confirm that all relevant data that support the findings of this study are included in the Letter and its Supplemental Material.

%\bibliographystyle{apsrev4-2}
%\bibliography{hybrid}


\clearpage
\newpage

\onecolumngrid

\section*{SUPPLEMENTAL MATERIAL}
\input{./SI_merge_contents.tex}

\end{document}

%% file: SI_merge_contents.tex
\setcounter{page}{1}
\setcounter{figure}{0}
\setcounter{table}{0}
\setcounter{section}{0}
\renewcommand{\thepage}{S\arabic{page}}
\renewcommand{\thesection}{S\arabic{section}}
\renewcommand{\thetable}{S\arabic{table}}
\renewcommand{\thefigure}{S\arabic{figure}}
\newcounter{SIfig}
\renewcommand{\theSIfig}{S\arabic{SIfig}}

%Section~\ref{Co-Ni} presents the additional results for the $A_4B_3$O$_{9}$ ($A$: La, Pr, Nd, Y; $B$: Co, Ni) layered oxides, while in Section~\ref{Fe} we present additional details for the La$A^\prime_3$Fe$_3$O$_{9}$ ($A^\prime$: Sr, Ca) systems.
Section~S1 presents the first-principles simulations details, while 
Sections~S2 and S3 contain additional structural details, electronic and magnetic properties of $A_4B_3$O$_{9}$ ($A$: La, Pr, Nd, Y; $B$ = Co, Ni) and La$A^\prime_3$Fe$_3$O$_9$ ($A^\prime$: Sr, Ca) layered oxides, respectively.

\section{S1. COMPUTATIONAL DETAILS}\label{methods}
%\section*{Methods}
First-principles calculations were performed within the density functional theory (DFT) framework using the projected augmented wave (PAW) method implemented in the VASP code~\cite{VASP1}, version 6.3.2. In order to accurately describe the equilibrium structures of the bulk $A_4B_3$O$_9$ oxides, PBEsol version of generalized gradient approximation (GGA) was chosen as the exchange-correlation functional~\cite{PBEsol}. PAW pseudopotentials (PBE, version 5.4)~\cite{pseudo} were used for all the calculations with the following valence configurations: $5s^25p^65d^16s^2$ (La), $5s^25p^65d^16s^2$ (Pr), $5s^25p^65d^16s^2$ (Nd), $4s^24p^25s^2$ (Sr), $3s^23p^24s^2$ (Ca), $3p^63d^74s^2$ (Co), $3p^63d^84s^2$ (Ni), $3p^63d^64s^2$ (Fe), and $2s^22p^4$ (O). Correlation effects were considered within the GGA+$U$ formalism introduced by Dudarev et al.~\cite{HubbardU} using an effective on-site Hubbard parameter $U_{\rm{eff}}$ = 5.0 eV, 5.15 eV and 4.0 eV for the $3d$ states of Co$^{2+}$, Ni$^{2+}$ and Fe$^{3+}$ ions, respectively~\cite{Hubbard_U_Co,Hubbard_U_Ni,Hubbard_U_Fe1,Hubbard_U_Fe2}. The $U_{\rm{eff}}$ values were varied within a reasonable range to check the stability of the ground states. Effect of Hund's parameter $J$ on weak ferromagnetic (wFM) moments was tested within the framework of Liechtenstein~\cite{Liechtenstein1995}
as implemented in VASP. Convergence tests were performed on a 64-atom $Cmcm$ unit cell of La$_4$Co$_3$O$_9$ which showed that a plane wave cutoff of 800 eV and $k$-mesh grid of $1 \times 5 \times 6$ in the full Brillouin zone (BZ) were sufficient to reach converged results. Depending on the symmetry of the structures, $k$-mesh grid was scaled accordingly. To check the numerical precision of our DFT calculations with respect to the energy differences between the polar and antipolar phases, we performed convergence tests on the polar $Cmc2_1$ and antipolar $Pmcn$ structures of La$_4$Co$_3$O$_9$ and Nd$_4$Co$_3$O$_9$. Our convergence tests with fixed lattice parameters (set at the values of the relaxed $Cmc2_1$ cell parameters) showed that a plane wave cutoff of 800 eV and $k$-mesh grid of $1 \times 5 \times 6$ ($1 \times 5 \times 5$) could resolve the energy difference between the polar and antipolar phases of La$_4$Co$_3$O$_9$ (Nd$_4$Co$_3$O$_9$) within $\sim$0.01 meV/f.u. Full relaxations were performed until the Hellmann–Feynman forces on each atom were less than 1 meV/\AA ~with an energy convergence criterion set at $10^{-9}$ eV. Spin-orbit coupling (SOC) effects were included self-consistently in the non-collinear calculations. We used the finite displacement method implemented in PHONOPY~\cite{Togo_2015} to calculate the phonon dispersions using a $1 \times 2 \times 2$ phonon supercell. Phonon spectra were computed for the primitive cell using the transformation matrix [0.5 -0.5 0.0, 0.5 0.5 0.0, 0.0 0.0 1.0]. We employed the Berry phase method~\cite{Vanderbilt1,Vanderbilt2} within VASP to compute the spontaneous electric polarization. The web-based ISOTROPY software suit~\cite{isotropy} was used for symmetry mode analyses, and visualizations of crystal structures and magnetic configurations were done by VESTA~\cite{VESTA}.  

\newpage
\section{S2. $A_4B_3\rm{O}_{9}$ ($A$: L\lowercase{a}, P\lowercase{r}, N\lowercase{d}, Y; $B$: C\lowercase{o}, N\lowercase{i}) layered oxides}\label{Co-Ni} 
\subsection{A. STRUCTURAL DETAILS}

\begin{table}[h]
  \setlength{\tabcolsep}{8.0pt}
  \caption{Fully optimized lattice parameters and magnetic moments of the different structural phases of the considered $A_{4}B_3$O$_{9}$ ($A$: La, Pr, Nd, Y; $B$: Co, Ni)  compounds calculated with G-AFM magnetic ordering using $U_{\rm{eff}}$ = 5.0 eV and 5.15 eV for the $3d$ states of Co and Ni, respectively. $\Delta E$ denotes the relative energy of different structural variants. Available experimental values are given in the parentheses for comparison. Note that hydration can lead to an expansion of the lattice parameter along the long axis~\cite{LaSr2016} and might explain the slight overestimation of the $a$-lattice parameter in Ref.~\cite{LaCo1998}. }
  \label{tab-lattice}
  \centering
  \begin{tabular}{|c|c|c|c c c c|c c|}
    \hline
    \hline\rule{0pt}{1.0\normalbaselineskip}
    Layered&Phase&$\Delta E$ &\multicolumn{4}{c|}{DFT-optimized lattice parameters}& \multicolumn{2}{c|}{Magnetic moments ($\mu_{\rm B}$)} \\\cline{4-7} \cline{8-9} \rule{0pt}{1.0\normalbaselineskip}
     oxide& &(meV/f.u.)& $a$ (\AA) & $b$ (\AA)& $c$ (\AA)& $V_{\rm{cell}}$ (\AA$^3$) & $\mu_{\rm{tetra}}$  & $\mu_{\rm{octa}}$  \\
     \hline \rule{0pt}{1.0\normalbaselineskip}
     &$Cmcm$&630.26&27.6366&5.6561&5.4374&849.95&2.545&2.660\\
     \rule{0pt}{1.0\normalbaselineskip}
    La$_{4}$Co$_3$O$_{9}$&$Cmc2_1$&0.42&28.0116&5.6812&5.3970&858.87&2.664&2.650\\
     \rule{0pt}{1.0\normalbaselineskip}
    &$Pmcn$&0.00&28.0122 &5.6808 &5.3978 &858.95 &2.664 &2.650 \\
    \rule{0pt}{1.0\normalbaselineskip}

    &&&(28.4600$^\dagger$)&(5.6467$^\dagger$) &(5.4356$^\dagger$)& (873.52$^\dagger$)&(2.6$^\dagger$)&(3.0$^\dagger$)\\
   \rule{0pt}{1.0\normalbaselineskip}
    
    &&&(27.5487$^\ddagger$)&(5.8568$^\ddagger$) &(5.6950$^\ddagger$)& (918.87$^\ddagger$)&&\\
    \hline \rule{0pt}{1.0\normalbaselineskip}
    
    &$Cmcm$&1087.92&27.3384&5.6392 &5.3930 &831.41 &2.666&2.655\\
     \rule{0pt}{1.0\normalbaselineskip}
    Pr$_{4}$Co$_3$O$_{9}$&$Cmc2_1$&0.11&27.7783&5.6749&5.3414&842.02&2.668&2.647\\
     \rule{0pt}{1.0\normalbaselineskip}
    &$Pmcn$&0.00&27.7796&5.6749&5.3420&842.14&2.668&2.647\\
    \hline \rule{0pt}{1.0\normalbaselineskip}

    &$Cmcm$&1266.50&27.9075&5.3850&5.2727&792.39&2.477&2.612\\
     \rule{0pt}{1.0\normalbaselineskip}
    Nd$_{4}$Co$_3$O$_{9}$&$Cmc2_1$&0.00&27.7716&5.5427&5.4997&846.56&2.666&2.672\\
     \rule{0pt}{1.0\normalbaselineskip}
    &$Pmcn$&3.20&27.7353&5.5943&5.4326&842.92&2.669&2.663\\
    \hline \rule{0pt}{1.0\normalbaselineskip}

    &$Cmcm$&2524.58&27.2707&5.2982&5.2050&752.04&2.488&2.630\\
     \rule{0pt}{1.0\normalbaselineskip}
    Y$_{4}$Co$_3$O$_{9}$&$Cmc2_1$&0.00&25.9609&5.3990&5.6600&793.33&2.684&2.694\\
     \rule{0pt}{1.0\normalbaselineskip}
    &$Pmcn$&35.10&26.0835&5.3795&5.6971&799.40&2.684&2.695\\
    \hline \rule{0pt}{1.0\normalbaselineskip}
     &$Cmcm$&552.09&27.5656&5.5510&5.4521&834.26&1.550&1.650\\
     \rule{0pt}{1.0\normalbaselineskip}
    La$_{4}$Ni$_3$O$_{9}$&$Cmc2_1$&0.00&28.0082&5.5801&5.3886&842.18&1.683&1.640\\
     \rule{0pt}{1.0\normalbaselineskip}
    &$Pmcn$&0.15&28.0105&5.5794&5.3886&842.14&1.683&1.640\\
    \hline \rule{0pt}{1.0\normalbaselineskip}
    &$Cmcm$&729.75&27.4044&5.5399&5.3886&818.10&1.542&1.646\\
     \rule{0pt}{1.0\normalbaselineskip}
    Pr$_{4}$Ni$_3$O$_{9}$&$Cmc2_1$&0.00&27.6193&5.6322&5.3250&828.35&1.690&1.636\\
     \rule{0pt}{1.0\normalbaselineskip}
    &$Pmcn$&0.32&27.8083&5.5761&5.3244&825.60&1.690&1.636\\
    \hline \rule{0pt}{1.0\normalbaselineskip}
    &$Cmcm$&761.68&27.1764&5.5330&5.3644&806.62&1.547&1.646\\
     \rule{0pt}{1.0\normalbaselineskip}
    Nd$_{4}$Ni$_3$O$_{9}$&$Cmc2_1$&0.00&27.3093&5.6398&5.2944&815.44&1.702&1.639\\
     \rule{0pt}{1.0\normalbaselineskip}
    &$Pmcn$&0.71&27.3126&5.6399&5.2933&815.38&1.702&1.639\\
    \hline \rule{0pt}{1.0\normalbaselineskip}
    &$Cmcm$&1827.87&26.3662&5.4469&5.3291&765.34&1.587&1.662\\
     \rule{0pt}{1.0\normalbaselineskip}
    Y$_{4}$Ni$_3$O$_{9}$&$Cmc2_1$&0.00&26.1242&5.3224&5.6156&780.82&1.703&1.670\\
     \rule{0pt}{1.0\normalbaselineskip}
    &$Pmcn$&3.70&26.2506&5.2961&5.6520&785.78&1.708&1.674\\
    \hline 
    \hline 
    \multicolumn{9}{l}{$\dagger$ denotes the experimental values measured in Ref.~\cite{LaCo1998}.} \\
    \multicolumn{9}{l}{$\ddagger$ denotes the experimental values measured in Ref.~\cite{Rahmani2022}.} \\
\end{tabular}
\end{table}
\clearpage

\subsection{B. ELECTRONIC PROPERTIES}
\begin{figure}[ht!]
\includegraphics[scale=1.316]{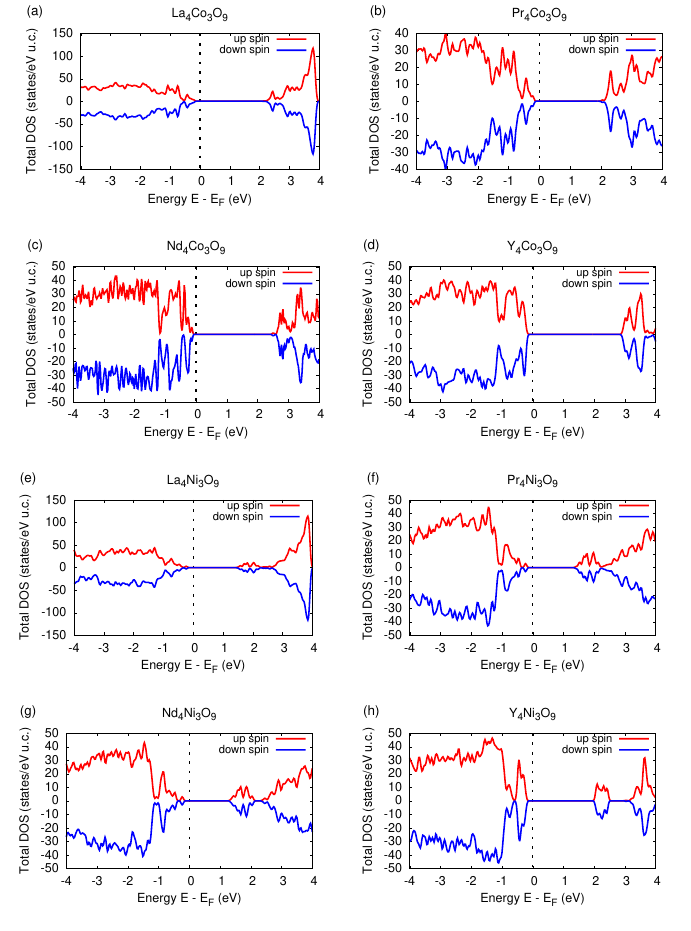}
    \caption{Total density of states (DOS) calculated for the lowest energy phases of $A_{4}B_3$O$_{9}$ ($A$: La, Pr, Nd, Y; $B$: Co, Ni) compounds (see Table~\ref{tab-lattice}) showing the insulating nature of the considered layered oxides. G-AFM magnetic order is considered with $U_{\rm{eff}}$ = 5.0 eV and 5.15 eV for the $3d$ states of Co and Ni, respectively.  }
\refstepcounter{SIfig}\label{dos}
\end{figure}
\clearpage

\subsection{C. GROUND STATE MAGNETIC CONFIGURATIONS}\label{magconfigs}
\begin{figure}[ht!]
\includegraphics[scale=0.53]{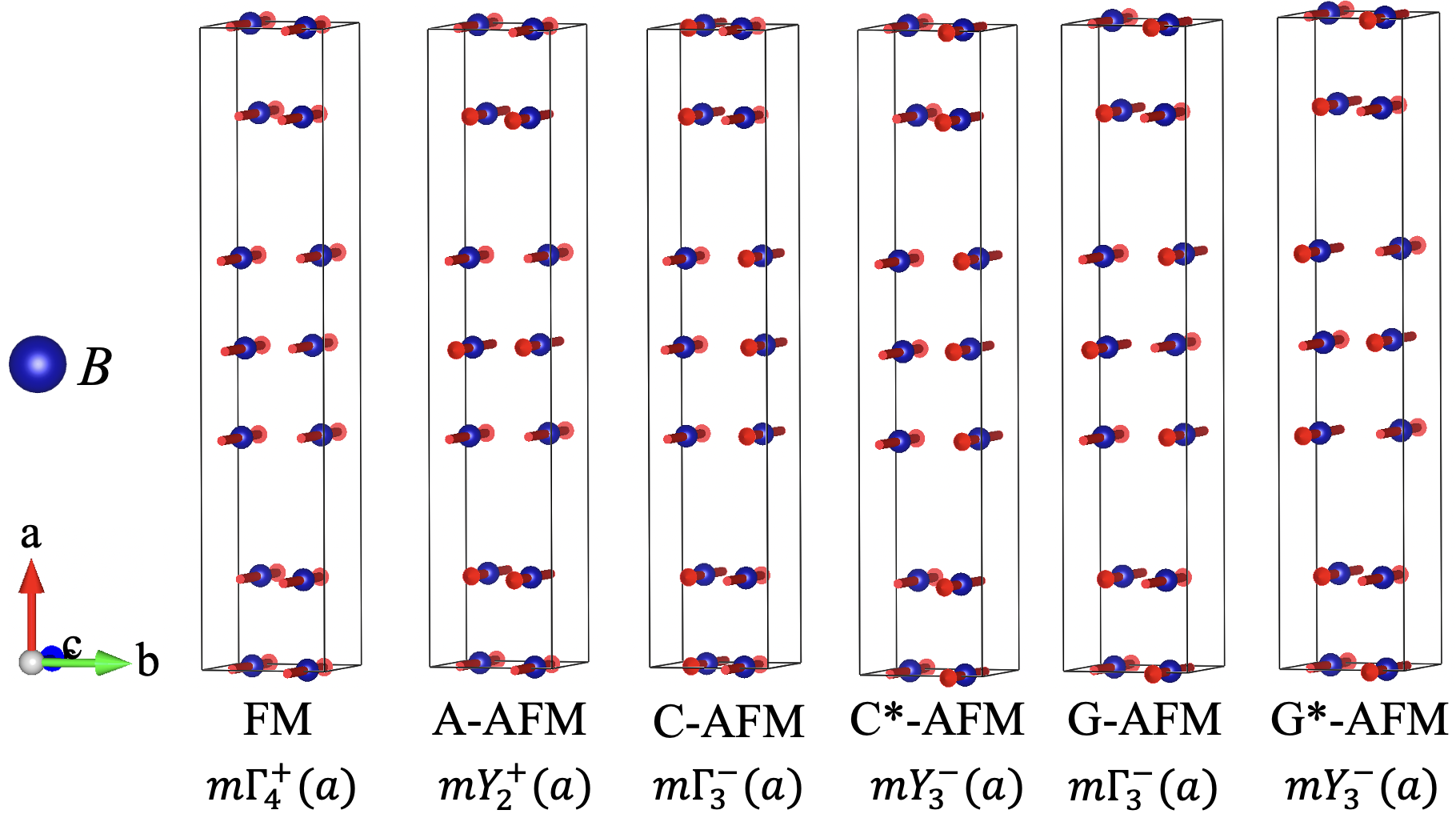}
    \caption{Collinear spin configurations of the $B$-site cations considered for calculating the ground state magnetic structure of the $A_4B_3$O$_{9}$ layered oxides. The corresponding magnetic irreps are also given. Spin magnetic moments are denoted by red arrows.}
\refstepcounter{SIfig}\label{magnstruct}
\end{figure}

Considering four initial spin arrangements of the $B$-site cations along the $c$-direction, namely the ferromagnetic (FM), A-type antiferromagnetic (A-AFM), C-type antiferromagnetic (C-AFM) and G-type antiferromagnetic (G-AFM), shown in Fig.~\ref{magnstruct}, we calculate the ground state magnetic configuration for the lowest energy nuclear structure of the considered layered oxides given in Table~\ref{tab-lattice}. Furthermore, we identify C$^*$- and G$^*$-AFM spin configurations, shown in Fig.~\ref{magnstruct}, which are very close in energy to the C- and G-AFM configurations, respectively (see Table~\ref{tab-mag}). The very small energy difference can be attributed to the very weak coupling between the two triple-layered perovskite blocks within the unit cell (see Table~\ref{tab-exchange}). Note that C-AFM, C$^*$-AFM, G-AFM and G$^*$-AFM configurations have identical energies in the parent $I4/mmm$ phase of the oxidized $A_4B_3$O$_{10}$ compounds, but possess different energies in presence of oxygen vacancy-induced octahedral rotations and hence should be treated separately for the symmetry analyses of $A_4B_3O_9$ layered oxides. Similarly, it is possible to have F$^*$- and A$^*$-AFM configurations by changing the relative coupling between the perovskite blocks in the FM and A-AFM spin structures. However, since FM and A-AFM orderings are found to be higher in energy in the considered $A_4B_3$O$_{9}$ oxides (see Table~\ref{tab-mag}), they are not discussed further. 
\clearpage

\clearpage
\begin{table}[ht!]
  \setlength{\tabcolsep}{2.5pt}
  \caption{Ground state (G.S.) magnetic configuration of the $A_4B_3$O$_{9}$ ($A$: La, Pr, Nd, Y; $B$: Co, Ni) layered oxides calculated for the lowest energy phases given in Table~\ref{tab-lattice}. $U_{\rm{eff}}$ = 5.0 eV and 5.15 eV are used for the $3d$ states of Co and Ni, respectively.  Note that Nd$_{4}$Co$_3$O$_{9}$ and $A_4$Ni$_3$O$_{9}$ possess wFM in the polar $Cmc2_1$ structure required for the $\Gamma$-point switching scheme described in Fig.~1(b) of the main text. We have used the non-standard $bca$ setting of the magnetic space group $Pn^\prime m^\prime a$ ($Pm^\prime c n^\prime$) to give the long axis along $a$ for consistency with the rest of the paper. While the energy difference between G-(C-) and G$^*$-(C$^*$-)AFM configurations is very small, we believe differences down to $\sim$0.02 meV/f.u. are real since this appears to be the level of agreement between G-G$^*$ vs. C-C$^*$ AFM orderings (which are expected to be the same under the assumption of the 5 primary $J$ exchange paths mentioned in Section~S2 J).  }
  \label{tab-mag}
  \centering
  \begin{tabular}{|c|c|c c c c c c|c|c|c|c|}
    \hline
    \hline\rule{0pt}{1.2\normalbaselineskip}
    Layered&Phase&\multicolumn{6}{c|}{Relative energy in meV/f.u. }&G.S. magnetic&Polar&Magnetic modes&wFM\\\cline{3-8}  \rule{0pt}{1.2\normalbaselineskip}
    oxide&& FM & A-AFM & C-AFM& C$^*$-AFM & G-AFM&G$^*$-AFM&space group&mode&present&  \\
     \hline \rule{0pt}{1.2\normalbaselineskip} 
    %&$Cmcm$&524.82&179.11 &20.89 &0.00 &0.09&$Pm^\prime cn^\prime$&\xmark&\cmark\\
     %\rule{0pt}{1.5\normalbaselineskip}
    La$_{4}$Co$_3$O$_{9}$&$Pmcn$&174.14&137.83 &32.39&32.41 &0.001 &0.00&$Pm^\prime cn^\prime$&\xmark&$m\Gamma^+_4(a)$, $mY^-_3(a)$ &\cmark \\
    \hline \rule{0pt}{1.5\normalbaselineskip} 

    %Pr$_{4}$Co$_3$O$_{9}$&$Cmcm$&597.02 &135.63 &46.58 &0.00&0.07&$Pm^\prime cn^\prime$&\xmark&\cmark \\
    %\rule{0pt}{1.5\normalbaselineskip} 
    Pr$_{4}$Co$_3$O$_{9}$&$Pmcn$&176.80 &138.71 &34.68&34.63 &0.06&0.00&$Pm^\prime cn^\prime$&\xmark&$m\Gamma^+_4(a)$, $mY^-_3(a)$&\cmark \\
    \hline \rule{0pt}{1.5\normalbaselineskip} 
    %Nd$_{4}$Co$_3$O$_{9}$&$Cmcm$&447.30 &401.50 & &0.00 &0.17&$Cm^\prime c2^\prime_1$&\cmark&\cmark \\
    %\rule{0pt}{1.5\normalbaselineskip}     
    Nd$_{4}$Co$_3$O$_{9}$&$Cmc2_1$&172.90 &136.88 &31.57&31.71 &0.00 &0.13&$Cm^\prime c2^\prime_1$&\cmark&$m\Gamma^+_4(a)$, $m\Gamma^-_3(a)$&\cmark\\
    \hline \rule{0pt}{1.2\normalbaselineskip} 
    %La$_{4}$Ni$_3$O$_{9}$&$Cmcm$&& & &0.00 &0.35&&& \\
   %\rule{0pt}{1.5\normalbaselineskip}     
   Y$_{4}$Co$_3$O$_{9}$&$Cmc2_1$& 135.29&106.08 &24.62&24.66 &0.00 &0.06&$Cm^\prime c2^\prime_1$&\cmark&$m\Gamma^+_4(a)$, $m\Gamma^-_3(a)$&\cmark\\
    \hline \rule{0pt}{1.2\normalbaselineskip} 
   La$_{4}$Ni$_3$O$_{9}$&$Cmc2_1$&267.72 &286.44 &0.00&0.28 &21.54& 21.78&$Cm^\prime c 2^\prime_1$&\cmark&$m\Gamma^+_4(a)$, $m\Gamma^-_3(a)$&\cmark \\
    \hline \rule{0pt}{1.2\normalbaselineskip}
    %Pr$_{4}$Ni$_3$O$_{9}$&$Cmcm$& & & &0.00 &0.34&&& \\
     %\rule{0pt}{1.5\normalbaselineskip}     
     Pr$_{4}$Ni$_3$O$_{9}$&$Cmc2_1$&273.16 &292.54&0.00 &0.18 &21.58&21.74&$Cm^\prime c 2^\prime_1$&\cmark&$m\Gamma^+_4(a)$, $m\Gamma^-_3(a)$&\cmark \\
    \hline \rule{0pt}{1.2\normalbaselineskip} 
    %Nd$_{4}$Ni$_3$O$_{9}$&$Cmcm$& & & &0.00 &0.37&&& \\
    %\rule{0pt}{1.5\normalbaselineskip}        
    Nd$_{4}$Ni$_3$O$_{9}$&$Cmc2_1$&259.64 &278.81 &0.00&0.15 &20.55 &20.69&$Cm^\prime c 2^\prime_1$&\cmark&$m\Gamma^+_4(a)$, $m\Gamma^-_3(a)$&\cmark \\
    \hline \rule{0pt}{1.2\normalbaselineskip} 

    Y$_{4}$Ni$_3$O$_{9}$&$Cmc2_1$&119.63 &144.31 &0.00&0.06 &38.89 &38.97&$Cm^\prime c 2^\prime_1$&\cmark&$m\Gamma^+_4(a)$, $m\Gamma^-_3(a)$&\cmark \\
    \hline 
    \hline
\end{tabular}
\end{table}

\begin{table}[ht!]
  \setlength{\tabcolsep}{6.0pt}
  \caption{Relative energies of the magnetic easy axis directions calculated for the lowest energy phases of Nd$_4$Co$_3$O$_{9}$ and Nd$_4$Ni$_3$O$_{9}$ as representatives of the Co- and Ni-series layered oxides, respectively. G-AFM spin ordering is considered for the noncollinear calculations including spin-orbit coupling (SOC). $U_{\rm{eff}}$ = 5.0 eV and 5.15 eV are used for the $3d$ states of Co and Ni, respectively.}
  \label{tab-easydir}
  \centering
  \begin{tabular}{|c|c|c c c|}
    \hline
    \hline\rule{0pt}{1.2\normalbaselineskip}
    Layered&Phase&\multicolumn{3}{c|}{Relative energy in meV/f.u. }\\\cline{3-5}  \rule{0pt}{1.2\normalbaselineskip}
     oxide& & along $a$ &along $b$ & along $c$  \\
     \hline \rule{0pt}{1.2\normalbaselineskip}   
    Nd$_{4}$Co$_3$O$_{9}$&$Cmc2_1$& 5.81&0.95 &0.00 \\
    \rule{0pt}{1.5\normalbaselineskip}     
    Nd$_{4}$Ni$_3$O$_{9}$&$Cmc2_1$& 2.16&1.34 &0.00  \\
    \hline 
    \hline
\end{tabular}
\end{table}

Note that the G$^*$-AFM magnetic configuration, which has been reported to be the ground state for the Fe-series compounds~\cite{LaSr2019}, does not allow for a wFM mode in the $Cmc2_1$ phase by symmetry but is very close in energy to the G-AFM magnetic structure of $A_4B_3$O$_9$ oxides. The small energy difference between the G- and  G$^*$-AFM spin orderings might lead to stacking faults in the magnetic structures of the Co-series layered oxides~\cite{Yamada1989,emma2014}. However, if G- vs. G$^*$-AFM ordering in Nd$_4$Co$_3$O$_9$ becomes an issue experimentally (i.e., there are domains of each), one could try to stabilize the G-AFM magnetic ordering by cooling through the phase transition with electric and magnetic fields. Furthermore, substitutional doping at the transition metal site might also allow the interlayer interactions to be tuned, stabilizing one magnetic structure over the other~\cite{Porter2022}.

\clearpage
\subsection{D. ATOMISTISTIC DETAILS BEFORE AND AFTER POLARIZATION REVERSAL} \label {atomic_structures}

\begin{figure}[ht!]
\includegraphics[scale=0.5]{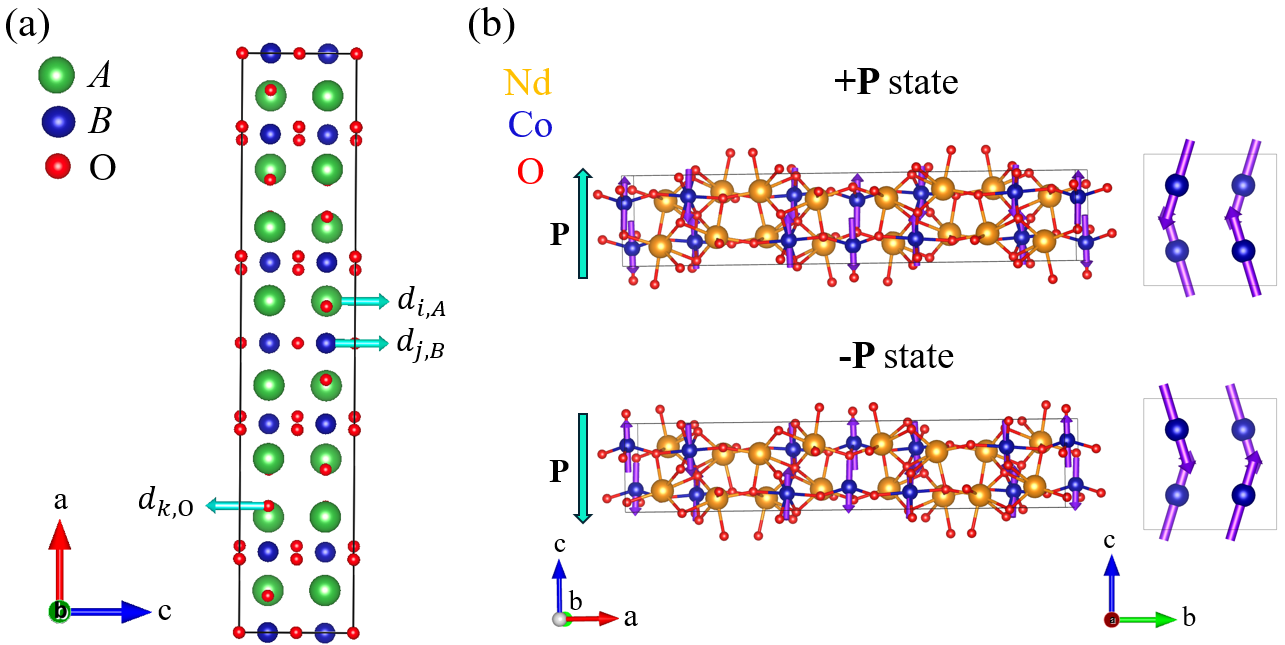}
    \caption{(a) Polar displacements of the $i$-th $A$-site atom, $j$-th $B$-site atom, and $k$-th O atom in the $Cmc2_1$ phase of $A_4B_3$O$_9$ layered oxides away from their equilibrium positions in the nonpolar $Cmcm$ phase. (b) Atomic structures in the $+\mathbf{P}$ and $-\mathbf{P}$ states of Nd$_4$Co$_3$O$_9$. Purple arrows indicate the spin magnetic moments on the Co$^{2+}$ ions. Magnetic moments on the octahedral sites (magnitude of the canting is slightly exaggerated), projected onto the $bc$-plane, illustrate the changes in the spin canting directions on polarization reversal.}
\label{atomicstruct}
\end{figure}

%DZYALOSHINSKII-MORIYA
Tetrahedral rotations around the out-of-plane axis in the $Cmc2_1$ phase leads to a finite net polarization ($\mathbf{P}$) along the [001] direction. %arising from the the displacement of cations away from the center of each tetrahedron. 
Fig.~\ref{atomicstruct}(a) shows the polar displacements of the atoms away from their equilibrium positions in the nonpolar $Cmcm$ phase, where $d_{i, A}$, $d_{j, B}$, and $d_{k, \rm{O}}$ denote the displacements of the $i$-th $A$-site atom, $j$-th $B$-site atom, and $k$-th O atom, respectively. The polar amplitude in Table I of the main text, $Q_{\Gamma^-_2}$, is defined as the root-summed-squared displacement of all the atoms in the $Cmc2_1$ phase influenced by the polar mode ($A_p$ value as defined in ISODISTORT~\cite{Campbell1,Campbell2}).
\begin{equation}
    Q_{\Gamma^-_2} = \sqrt{\sum^{N_A}_{i = 1} (d_{i, A})^2 + \sum^{N_B}_{j = 1} (d_{j, B})^2 + \sum^{N_{\rm{O}}}_{k = 1} (d_{k, \rm{O}})^2}
\end{equation}
where $N_A, N_B, N_{\rm{O}}$ are, respectively, the total number of $A$-site atoms, $B$-site atoms, and O atoms in the conventional unit cell. It is important to note that, due to the well-known issue of origin ambiguity in polar structures, the relative displacements of the cations and anions shown in Fig.~\ref{atomicstruct}(a) are presented for schematic purposes only. In our work, we have chosen the origins set by ISODISTORT~\cite{Campbell1,Campbell2}.  

%The local electric dipoles from the $A$- and $B$-site atoms along [001] are shown in Fig.~\ref{atomicstruct}(b-c). In the $Cmc2_1$ phase, the polar mode couples to the antiferromagnetically ordered (C- and G-AFM) $B$-site spins along [001] (as shown in Fig.~\ref{magnstruct}) to induce wFM moments via the trilinear coupling in Fig.~1(b) of the main text. Presence of wFM leads to a canting of the collinear AFM spins (described by AFM vector $L_c$) and gives rise to a net magnetization $M_b$ along [010] through the Dzyaloshinskii-Moriya (DM) interaction: $E_{\rm{DM}} = -\mathbf{D} \cdot (\mathbf{L} \times \mathbf{M})$~\cite{DMI1958,Moriya1960}. Then by symmetry, the DM vector $\mathbf{D}$ points along the [100] direction in the $+\mathbf{P}$ state. When the direction of polarization changes from $+\mathbf{P}$ to $-\mathbf{P}$, the small wFM moments along [010] spontaneously change directions leading to a 180$\degree$ reversal of the net magnetization (from $M_b$ to $+\mathbf{M}$) through the $\Gamma$-point scheme, accompanied by a reversal of $\mathbf{D}$ and the spin canting directions, since the primary AFM vector $L_c$ is less likely to switch due to magnetocrystalline anisotropy (see Section~\ref{AFM_switching} for details).}

Atomistic details illustrating the changes in spin canting directions before and after polarization reversal in the $Cmc2_1$ ground state of Nd$_4$Co$_3$O$_9$ are shown in Fig.~\ref{atomicstruct}(b). In the $+\mathbf{P}$ state, the atoms develop a finite polarization along the $c$-direction ($P_c$). The Dzyaloshinskii–Moriya interaction (DMI) induces a finite canting of the G-AFM moments aligned along the $c$-axis ($L_c$), resulting in a net magnetization along the $-b$-direction ($M_b$).

\clearpage
\subsection{E. EFFECT OF POLARIZATION REVERSAL ON DZYALOSHINSKII–MORIYA INTERACTION}

Table~\ref{symmetry} demonstrates that a trilinear coupling term of the form $P_cL_cM_b$ remains invariant under the combined actions of the symmetry operations of the parent $Cmcm$ phase and time-reversal symmetry, consistent with the $\Gamma$-point scheme described in the main text. More specifically, this coupling gives the following contribution to the expansion of the free energy around the high symmetry phase:
\begin{equation}
    \mathcal{F}(P_c, L_c, M_b) = \gamma P_cL_cM_b,
\label{trilinear}
\end{equation}
where $\gamma$ is the expansion coefficient. 
\begin{table}[ht!]
  \setlength{\tabcolsep}{6.0pt}
  \caption{Changes in the directions of polarization ($P_c$), AFM moments ($L_c$), and net magnetization ($M_b$) under the symmetry operations of the parent $Cmcm$ phase and time-reversal symmetry ($\mathbbm{1}^\prime$). $\mathbbm{1}$ and $\bar{\mathbbm{1}}$ stand for the identity and inversion operations, respectively. 2 denotes a twofold rotation around the crystallographic axes. From the transformations in the Table, it follows that the product $P_c L_c M_b$ is invariant.}
  \label{symmetry}
  \centering
  \begin{tabular}{|c|c|c|c|c|c|}
    \hline
    \hline \rule{0pt}{1.2\normalbaselineskip}
     $\mathbbm{1}$ & $\bar{\mathbbm{1}}$& $2_x$ & $2_y$ & $2_z$ & $\mathbbm{1}^\prime$\\ \rule{0pt}{1.2\normalbaselineskip}
    &&$+ (0, 0, \frac{1}{2})$& $+ (0, 0, \frac{1}{2})$ & $+ (0, 0, \frac{1}{2})$ &\\ 
    \hline \rule{0pt}{1.2\normalbaselineskip}
  $L_c$& $-L_c$ & $L_c$ & $-L_c$  & $-L_c$  & $-L_c$   \\
   \hline \rule{0pt}{1.2\normalbaselineskip}
  $P_c$&  $-P_c$& $-P_c$ & $-P_c$ & $P_c$ & $P_c$\\
   \hline \rule{0pt}{1.2\normalbaselineskip}
  $M_b$& $M_b$ & $-M_b$ & $M_b$ & $-M_b$ & $-M_b$\\
    \hline 
    \hline
\end{tabular}
\end{table}

It is known that the Dzyaloshinskii–Moriya (DM) vector ($\mathbf{D}$) forms a right-handed system with $\mathbf{L}$ and $\mathbf{M}$ i.e., the energy term associated with the DMI in Nd$_4$Co$_3$O$_9$ takes the form,  $E_{\rm{DMI}} = D_aL_cM_b$~\cite{DMI1958,Moriya1960,Ederer2005,Ederer2006}. Then Eq.~(\ref{trilinear}) implies that the direction of the DM vector is correlated with the direction of polarization: $D_a \propto \gamma P_c$. Note that there is no term independent of $P_c$ in $D_a$, since the DMI arises together with the polar mode in the $Cmc2_1$ structure. Consequently, the reversal of polarization along $c$ would lead to the reversal of the DM vector pointing along the $a$-direction (orthogonal to the mirror planes containing the spin moments).  It would be accompanied by a reversal of the spin canting directions, as shown in Fig.~\ref{atomicstruct}(b), since the switching of the primary G-AFM ordering is energetically less favorable (see Section~S2 I for details).

\clearpage
\subsection{F. PHONON SPECTRA AND METASTABLE PHASES} \label {phonons}

\begin{figure}[ht!]
\includegraphics[scale=0.56]{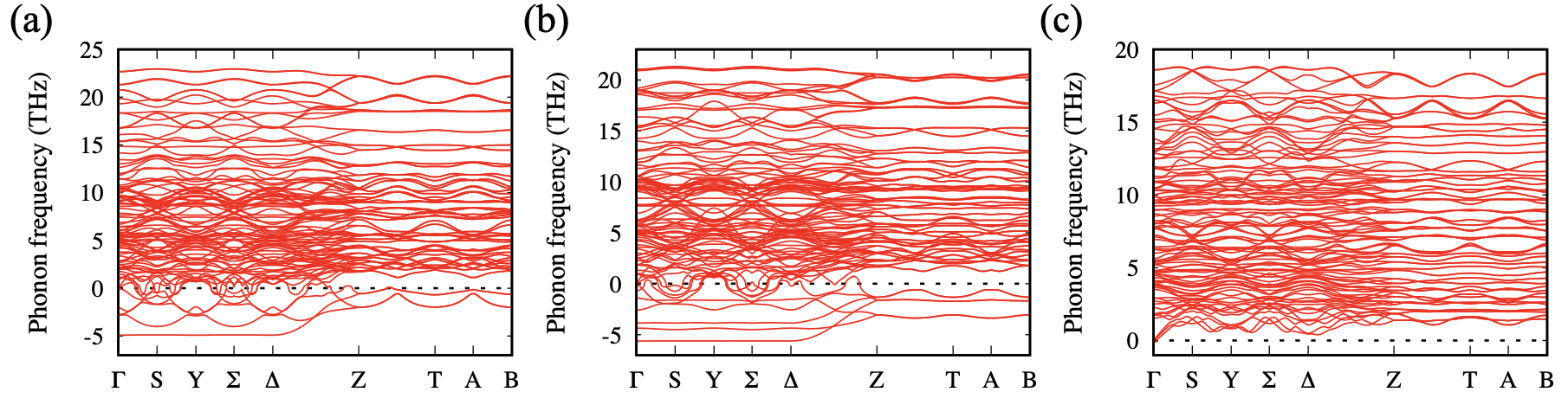}
    \caption{Phonon spectra along the $\Gamma (0, 0, 0)-S(0, 0.5, 0)-\Sigma (0.25, 0.25, 0)-Y(0.5, 0.5, 0)-\Delta(0.75, 0.25, 0) -Z(0, 0, 0.5)-T(0.5, 0.5, 0.5)-A(0.25, 0.25, 0.5)-B(0.75, 0.25, 0.5)$ direction in the BZ (in the primitive basis) calculated with G-AFM magnetic order for the high symmetry $Cmcm$ phase of (a) Nd$_{4}$Co$_3$O$_{9}$ and (b) Nd$_{4}$Ni$_3$O$_{9}$ as representatives of the Co- and Ni-series $A_4B_3$O$_{9}$ layered oxides, respectively. $U_{\rm{eff}}$ = 5.0 eV and 5.15 eV are used for the $3d$ states of Co and Ni, respectively. As seen, all the Co- and Ni-series compounds have a strongly unstable flat phonon branch along the $\Gamma-S-\Sigma-Y-\Delta$ direction in the BZ associated with the tetrahedral chain ordering distortions. Panel (c) shows the phonon bands for the polar $Cmc2_1$ phase of Nd$_{4}$Co$_3$O$_{9}$. Absence of imaginary phonon frequencies indicates the dynamical stability of this polar phase. }
\label{phonon}
\end{figure}
Phonon spectra in Fig.~\ref{phonon} show that the paraelectric $Cmcm$ phase of $A_4B_3$O$_{9}$ layered oxides is unstable against the tetrahedral chain ordering distortions. The phonon spectra of all the considered $A_{4}B_3$O$_{9}$ systems show qualitatively similar features with a strongly unstable flat phonon branch along the $\Gamma-$S$-\Sigma-$Y$-\Delta$ direction in the BZ related to the tetrahedral chain ordering distortions. Presence of a flat phonon band indicates
that the different structural variants derived from these instabilities will be close in energy (see Table~\ref{tab-sup}). The flat phonon band may have further implications, such as the induction of robust yet independently reversible electric polarization in each layer as observed in hafnium dioxide~\cite{hafnia} and is worthy of future investigation, potentially in collaboration with experimental studies.  

\begin{figure}[ht!]
\includegraphics[scale=0.5]{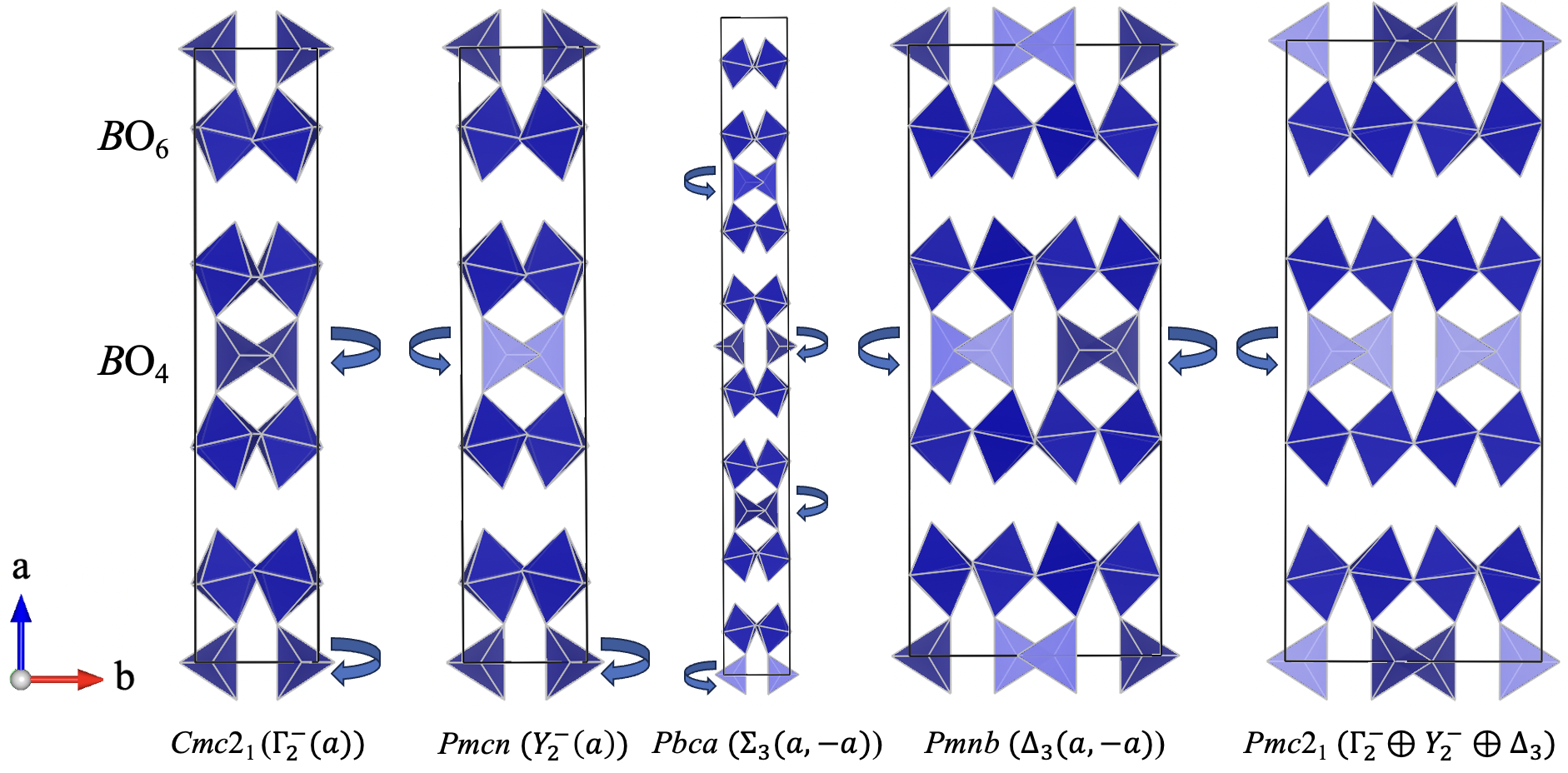}
    \caption{Different structural phases of $A_{4}B_3$O$_{9}$ layered oxides arising from cooperative rotations of the tetrahedral units.}
\label{superstruct}
\end{figure}

Therefore, in order to check the stability of the ground state polar $Cmc2_1$ structure of Nd$_{4}B_3$O$_{9}$ ($B$: Co, Ni) (Nd$_{4}$Ni$_3$O$_{9}$ as a representative of the Ni-series compounds, see Table~\ref{tab-lattice}) with non-trivial ME effect, we condense in the phonon instabilities at the zone-center and zone-boundary points which result in a number of structural variants with different space group symmetries as shown in Fig.~\ref{superstruct}. 

As mentioned in the main text, when all the tetrahedra rotate in the same direction, the dipole moments from each layer add up, resulting in a polar structure in the $Cmc2_1$ space group (no. 36) which is analogous to the $Ima2$ phase of brownmillerites (BMs)~\cite{Bellaiche2019} and $Pmc2_1$ phase of Grenier oxides~\cite{Shin2023}. The $Cmc2_1$ polar phase of the $A_4B_3$O$_9$ layered oxides is associated with a polar mode transforming as the $\Gamma^-_2(a)$ irrep of the parent $Cmcm$ phase. 
On the other hand, if the tetrahedra rotate in opposite senses in successive layers, the dipole moments from each layer cancel out forming an antipolar structure. This antipolar structure is related to the $Cmcm$ phase by the $Y^-_2(a)$ irrep which reduces the symmetry from $Cmcm$ to $Pmcn$ (no. 62). Again, this model is analogous to the $Pnma$ and $Pbcm$ models observed in BMs~\cite{Bellaiche2019} and Grenier phases~\cite{Shin2023}, respectively. Note that we have used the non-standard $bca$ setting of the space group $Pnma$
($Pmcn$) to give the long axis along $a$ for consistency with the rest of the paper. 

Different intralayer and interlayer tetrahedral twisting patterns can further lead to a variety of other distinct phases resulting in superstructures with longer periods.
An example of a superstructure with intralayer switching of tetrahedral rotation patterns described by $Pmnb$ symmetry (no. 62) is shown in Fig.~\ref{superstruct}, which is associated with the $\Delta_3(a,-a)$ irrep of the parent phase and similar to the $Pbcm$ phase of the BMs~\cite{Bellaiche2019}. On the other hand, more complex interlayer rotation patterns corresponding to the $\Sigma_3(a, -a)$ irrep, shown in Fig.~\ref{superstruct}, leads to a different antipolar phase in the $Pbca$ space group. 
We have identified another novel polar phase with $Pmc2_1$ symmetry which contributes an energy
term of the form $Q^2_{\Delta_3}Q_{\Gamma^-_2}Q_{Y^-_2}$
in the free energy
expansion of the high symmetry phase. This kind of novel ferroelectric phase with a quadratic-bilinear coupling has also been observed in BM oxides~\cite{Tian2018}. 

Relative energies of the fully relaxed metastable structures are listed in Table~\ref{tab-sup}. Note that we have limited our simulations to the 128 atom-superstructures
to keep the computations tractable.

\begin{table}[h]
  \setlength{\tabcolsep}{4.0pt}
  \caption{Relative energies of the different metastable phases of Nd$_{4}B_3$O$_{9}$ ($B$: Co, Ni) arising from the relative ordering of the tetrahedral twisting patterns shown in Fig.~\ref{superstruct}. We have fully relaxed the structures with G-AFM spin order using $U_{\rm{eff}}$ = 5.0 eV and 5.15 eV for the $3d$ states of Co and Ni, respectively. Note that the polar $Cmc2_1$ phase is the ground state of Nd$_{4}$Co$_3$O$_{9}$ which allows for the $\Gamma$-point switching scheme described in Fig.~1(b) of the main text. }
  \label{tab-sup}
  \centering
  \begin{tabular}{|c|c|c| c |c |}
    \hline
    \hline\rule{0pt}{1.2\normalbaselineskip}
    Phase&Irrep(s) of primary &Irrep(s) of secondary &\multicolumn{2}{c|}{Relative energy in meV/f.u. }\\\cline{4-5}  \rule{0pt}{1.5\normalbaselineskip}
     &order parameter(s)&order parameter(s)&Nd$_{4}$Co$_3$O$_{9}$&Nd$_{4}$Ni$_3$O$_{9}$  \\
     \hline \rule{0pt}{1.5\normalbaselineskip}   
    %&$Cmcm$&524.82&179.11 &20.89 &0.00 &0.09&$Pm^\prime cn^\prime$&\xmark&\cmark\\
     %\rule{0pt}{1.5\normalbaselineskip}
    $Cmc2_1$&$\Gamma^-_2(a)$&$\Gamma^+_1(a)$&0.00&8.55 \\
     \hline \rule{0pt}{1.5\normalbaselineskip} 
    $Pmcn$&$Y^-_2(a)$&$\Gamma^+_1(a)$&3.20&9.28 \\
    \hline \rule{0pt}{1.5\normalbaselineskip}
    $Pmab$&$\Sigma_3(0, a)$&$\Gamma^+_1(a)$&209.92&261.00 \\
      \rule{0pt}{1.5\normalbaselineskip}
     &&$Y^+_1(a)$&& \\
     \hline \rule{0pt}{1.5\normalbaselineskip}
     $Pbca$&$\Sigma_3(a, -a)$&$\Gamma^+_1(a)$&5.74&8.95 \\
     \rule{0pt}{1.5\normalbaselineskip}
     &&$Y^-_3(a)$&& \\
     \hline \rule{0pt}{1.5\normalbaselineskip}
     $Pmcb$&$\Delta_3(a, 0)$&$\Gamma^+_1(a)$&433.73&425.99 \\
      \rule{0pt}{1.5\normalbaselineskip}
     &&$Y^+_1(a)$&& \\
     \hline \rule{0pt}{1.5\normalbaselineskip}
     $Pmnb$&$\Delta_3(a, -a)$&$\Gamma^+_1(a)$&9.60&0.00 \\
      \rule{0pt}{1.5\normalbaselineskip}
     &&$Y^-_4(a)$&& \\
     \hline \rule{0pt}{1.5\normalbaselineskip}
     $Pmc2_1$&$\Gamma^-_2(a)$&$\Gamma^+_1(a)$&9.84&4.41 \\
      \rule{0pt}{1.5\normalbaselineskip}
     &$Y^-_2(a)$&$\Delta_1(a, 0)$&& \\
      \rule{0pt}{1.5\normalbaselineskip}
     &$\Delta_3(0, a)$&$Y^+_1(a)$&& \\
     \hline \rule{0pt}{1.5\normalbaselineskip}
     %$Pmc2_1$&$\Delta_3(a, b)$&$\Gamma^+_1(a)$&$-$&0.00 \\
     %\rule{0pt}{1.5\normalbaselineskip}
     %&&$\Gamma^-_4(a)$&& \\
     % \rule{0pt}{1.5\normalbaselineskip}
     %&&$Y^+_1(a)$&& \\
      \rule{0pt}{1.5\normalbaselineskip}
     %&&$Y^-_4(a)$&& \\
     %\hline \rule{0pt}{1.5\normalbaselineskip}
     
    $Cc2m$&$Z_2(a, 0)$&$\Gamma^+_1(a)$&728.15&626.70 \\
      \rule{0pt}{1.5\normalbaselineskip}
     &&$\Gamma^-_4(a)$&& \\
     %\hline \rule{0pt}{1.5\normalbaselineskip}
     %$Cc$&$Z_2(a, b)$&$\Gamma^+_1(a)$&$-$&7.94 \\
     %\rule{0pt}{1.5\normalbaselineskip}
     %&&$\Gamma^+_3(a)$&& \\
     % \rule{0pt}{1.5\normalbaselineskip}
     %&&$\Gamma^-_2(a)$&& \\
      %\rule{0pt}{1.5\normalbaselineskip}
     %&&$\Gamma^-_4(a)$&& \\
    \hline 
    \hline
\end{tabular}
\end{table}
\clearpage

\subsection{G. EFFECT OF $U_{\rm{eff}}$ ON THE GROUND STATE OF Nd$_4$Co$_3$O$_9$}
In order to check the stability of the ground state magnetic configuration and crystal structure of Nd$_4$Co$_3$O$_9$, we consider two other extreme sets of $U_{\rm{eff}}$ values for the Co-$3d$ states as described in Table~\ref{tab-U} and find that polar $Cmc2_1$ phase with G-AFM spin ordering of the Co$^{2+}$ ions is stable against different Hubbard $U_{\rm{eff}}$ parameters within reasonable range. Very small value of $U_{\rm{eff}}$ is, however, found to favor the antipolar structure. 

\begin{table}[ht!]
  \setlength{\tabcolsep}{12.0pt}
  \caption{Stability of the polar $Cmc2_1$ phase of Nd$_4$Co$_3$O$_{9}$ as a function of the $U_{\rm{eff}}$ parameter. $\Delta E$ denotes the relative energies of the polar $Cmc2_1$ and antipolar $Pmcn$ phases. Relative energies of the closely lying G-AFM and G$^*$-AFM magnetic configurations are also given for the lowest energy phase in each case. As seen, polar $Cmc2_1$ phase with G-AFM spin ordering of the Co$^{2+}$ ions is stable against different $U_{\rm{eff}}$ parameters within reasonable range. While the energy difference between G- and G$^*$-AFM configurations is very small, we believe differences down to $\sim$0.02 meV/f.u. are real (see Table~\ref{tab-mag} for details).}
  \label{tab-U}
  \centering
  \begin{tabular}{|c|c|c|c|c|c|}
    \hline
    \hline\rule{0pt}{1.2\normalbaselineskip}    $U_{\rm{eff}}$&Phase &$\Delta E$&\multicolumn{2}{c|}{Relative energy in meV/f.u. }\\\cline{4-5}  \rule{0pt}{1.2\normalbaselineskip}
     (eV) & & (meV/f.u.) & G-AFM & G$^*$-AFM \\
     \hline \rule{0pt}{1.2\normalbaselineskip}   
    3.0&$Cmc2_1$& 0.00&0.00 &0.26 \\
    \rule{0pt}{1.5\normalbaselineskip}     
     &$Pmcn$&0.58 & & \\
    \hline \rule{0pt}{1.5\normalbaselineskip} 
    5.0&$Cmc2_1$& 0.00&0.00 &0.13 \\
    \rule{0pt}{1.5\normalbaselineskip}     
     &$Pmcn$& 3.20& & \\
    \hline \rule{0pt}{1.5\normalbaselineskip} 
    8.0&$Cmc2_1$& 0.00&0.00 &0.05 \\
    \rule{0pt}{1.5\normalbaselineskip}     
     &$Pmcn$& 7.28& & \\
    \hline 
    \hline
\end{tabular}
\end{table}

\subsection{H. EFFECT OF $J$ ON THE WEAK FERROMAGNETIC MOMENT OF Nd$_4$Co$_3$O$_{9}$}
\begin{table}[ht!]
  \setlength{\tabcolsep}{12.0pt}
  \caption{Effect of the Hund's parameter $J$ on the net wFM moment calculated with G-AFM magnetic configuration of Co$^{2+}$ ions in the polar $Cmc2_1$ ($+\mathbf{P}$ state) phase of Nd$_4$Co$_3$O$_{9}$.}
  \label{tab-J}
  \centering
  \begin{tabular}{|c|c|c|}
    \hline
    \hline\rule{0pt}{1.2\normalbaselineskip}    
    $U$ (eV) & $J$ (eV) &Net wFM moment ($\mu_{\rm B}$/u.c.)\\  
     \hline \rule{0pt}{1.2\normalbaselineskip}   
    5.0&0.0&-0.18 \\
    \rule{0pt}{1.5\normalbaselineskip}     
     5.5&0.5&-0.20 \\
    \rule{0pt}{1.5\normalbaselineskip} 
     6.0&1.0&-0.24  \\
    \hline 
    \hline
\end{tabular}
\end{table}

\clearpage
\subsection{I. SWITCHING OF MAGNETIZATION VS. ANTIFERROMAGNETIC ORDER IN Nd$_4$Co$_3$O$_{9}$ DURING POLARIZATION REVERSAL} \label{AFM_switching}

Reversal of magnetization with polarization through the magnetoelectric coupling schemes described in Fig.~1 of the main text requires that the AFM order parameter does not change sign under the application of an electric field. In the following, we demonstrate that it is the wFM mode, rather than the AFM mode, that should change sign during the reversal of the polar mode.

\begin{enumerate}
    \item One approach to understanding this is to compute the relative switching barriers for the wFM ($M_b$) and G-AFM ($L_c$) order parameters in Nd$_4$Co$_3$O$_{9}$  during the polarization ($P_c$) reversal process. 

Switching of the primary AFM mode would require the G-AFM moments in Nd$_4$Co$_3$O$_{9}$ (magnitude of $\sim$2.67 $\mu_{\rm B}$/Co$^{2+}$) to rotate by 180\degree ~through the magnetic hard directions. Since the G-AFM order parameter $m\Gamma^-_3(a)$ is one-dimensional, the AFM moments can only point along a certain crystallographic direction. As shown in Table~\ref{tab-easydir}, the energy costs associated with the rotation of the magnetic moments through the in-plane (0.95 meV/f.u.) and out-of-plane (5.81 meV/f.u.) magnetic hard axes are sizeable, implying that the magnetocrystalline anisotropy will hinder the rotation of the AFM moments, making the switching of the AFM mode energetically less favorable. It is worth noting that the strong uniaxial magnetic anisotropy of high spin Co$^{2+}$ ions in (pseudo) octahedral environment is consistent with previous experimental reports on related Co(II) complexes~\cite{Babkevich2010,LLORET2008,Pato2016,Tachiki1960}.

As for the energy barrier associated to a wFM reversal, we studied it by running the following first-principles calculations. We start with the ground state trio ($P_c$, $L_c$, $M_b$), and then switch only the direction of $P_c$ resulting in the ($-P_c$, $L_c$, $M_b$) state.   
%when we start our calculations in the state with an opposite sign of the wFM mode  (i.e., with a net magnetization $M_b$ corresponding to the $+\mathbf{P}$ state), 
We find that the magnetic moments spontaneously change their canting directions and return to the ($-P_c$, $L_c$, $-M_b$) state during the self-consistent spin relaxations with SOC, without falling into the ($-P_c$, $-L_c$, $M_b$) state. This indicates that the switching of the secondary wFM mode (magnitudes of $\sim$0.025 $\mu_{\rm B}$/Co$^{2+}_{\rm{octa}}$ and $\sim$0.006 $\mu_{\rm B}$/Co$^{2+}_{\rm{tetra}}$ ) with $P_c$ is spontaneous and has a \textit{zero} energy barrier.

 \item Our DFT results on the $Cmcm$ structure with the G-AFM spin ordering show that $M_b$ is an improper order parameter that would not exist in the absence of $P_c$. Similarly, it can be established that $M_b$ does not exist in the absence of the AFM order $L_c$ i.e., the very existence of $M_b$ depends on the trilinear coupling $P_cL_cM_b$ (see Eq.~(\ref{trilinear})). Hence, there is no energy minimum associated to $M_b$ by itself. By contrast, $L_c$ is a proper order parameter that lies within an energy minimum defined by anisotropy energies. Hence, $M_b \sim P_cL_c$ is the very \textit{origin} of $M_b$. If $P_c$ or $L_c$ changes sign, $M_b$ will automatically change sign as well. More importantly, $M_b$ can never drag $L_c$ through an $L_c = 0$ state and force an $L_c$-reversal since $M_b$ cannot exist unless $L_c$ is different from zero.

 \item On reversal of $P_c$, the \textit{force} $f_{M_b}$ experienced by the $M_b$ order parameter scales as $\sim P_cL_c$, while the \textit{force} $f_{L_c}$ experienced by the $L_c$ order parameter scales as $\sim P_cM_b$. 
 \begin{equation*}
     f_{M_b} = -\frac{\partial \mathcal{F}}{\partial M_b} \propto P_cL_c; \quad f_{L_c} = -\frac{\partial \mathcal{F}}{\partial L_c} \propto P_cM_b 
 \end{equation*}
 
 Consequently, the ratio of forces is given by $f_{M_b}/f_{L_c} \sim L_c/M_b$. Since $L_c \gg M_b$, it follows that $f_{M_b} \gg f_{L_c}$. This implies that the driving force for $M_b$ reversal is significantly larger (by approximately $2-3$ orders of magnitude) than the corresponding force for $L_c$ reversal.
\end{enumerate}
  
Thus our first-principles calculations predict the reversal of net magnetization with an electric field in Nd$_4$Co$_3$O$_{9}$ as the switching of the G-AFM order parameter is energetically less favorable compared to the switching of the wFM mode. Similar predictions of magnetization reversal and/or changes in the spin canting directions with polarization has been made in Refs.~\cite{Fennie2008,Ederer2005,Spaldin2006} and numerically demonstrated in Refs.~\cite{Yang2014,Xu2024}.

Experimentally, one could grow Nd$_4$Co$_3$O$_9$ films on a substrate with the same G-AFM magnetic structure. This setup would make the electric field switching of the AFM ordering energetically much less favorable, as it would require switching of the AFM ordering in both the film and the substrate, thereby effectively pinning the direction of the AFM order parameter.
%cooling through the phase transition in presence of electric and magnetic fields would pin the $P_c$ and wFM order parameters to align with the fields, which would in turn pin the AFM order parameter to the corresponding preferred direction (due to the trilinear coupling between $P_c$, wFM, and AFM order parameters).

\clearpage
\subsection{J. MAGNETIC EXCHANGE INTERACTION PARAMETERS AND POSSIBILITY OF ROOM-TEMPERATURE MAGNETISM IN Nd$_4$Co$_3$O$_{9}$}\label{mag-exchange}

\begin{figure}[ht!]
\includegraphics[scale=0.6]{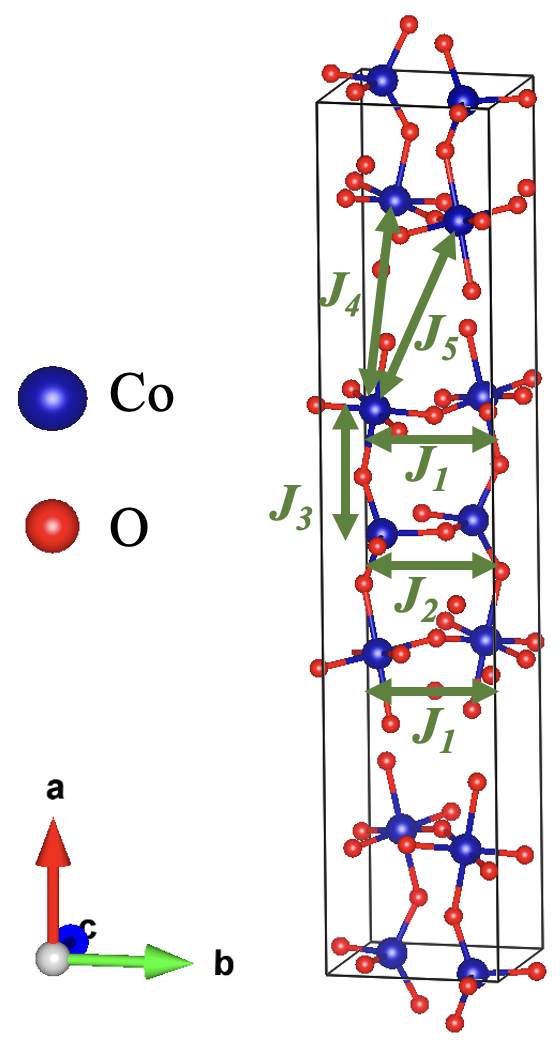}
    \caption{Crystal structure of Nd$_{4}$Co$_3$O$_{9}$ showing different nearest neighbor spin exchange interaction parameters considered in our simulations.}
\label{exchange}
\end{figure}

In order to investigate the possibility of room-temperature (RT) magnetism in Nd$_4$Co$_3$O$_{9}$, we calculate the nearest neighbor (NN) magnetic exchange interaction parameters by mapping the DFT-calculated total energies of different magnetic structures onto the classical Heisenberg model:
\begin{equation}
H = -\sum_{\langle ij \rangle} J_{ij} ~\mathbf{S}_i \cdot \mathbf{S}_j
\end{equation}
where $J_{ij}$ denotes the magnetic exchange interaction between the NN $i$ and $j$ sites. $J > 0$ indicates FM spin interactions, while $J < 0$ denotes AFM interactions between the spins. $J_{ij}$ runs from $J_1$ to $J_5$ as shown in Fig.~\ref{exchange}. $J_1$ and $J_2$ denote the in-plane interactions between the NN octahedral and tetrahedral sites, respectively. On the other hand, $J_3$ is the NN out-of-plane spin interaction within the perovskite blocks. $J_4$ and $J_5$ are slightly inequivalent interlayer exchange interactions which gives rise to G-(C-) vs. G$^*$-(C$^*$)AFM configurations. 

Considering the two perovskite blocks to be identical, we calculate the NN spin exchange parameters for the ground state polar $Cmc2_1$ phase of Nd$_4$Co$_3$O$_{9}$ and compare the $J^\prime$s with those computed for the experimentally observed antipolar $Pmcn$ phase of La$_4$Co$_3$O$_{9}$.  As seen from Table~\ref{tab-exchange},  the values of the corresponding $J^\prime$s are almost identical. Therefore, in analogy with the experimental RT long-range G-AFM ordering of Co$^{2+}$ ions in La$_4$Co$_3$O$_{9}$~\cite{LaCo1998}, we predict RT G-AFM magnetic ordering in the ground state polar $Cmc2_1$ phase of Nd$_4$Co$_3$O$_{9}$. Hence, we put forward Nd$_4$Co$_3$O$_{9}$ as an ideal candidate to observe the electric field reversal of magnetization at RT with immense technological importance in the next generation memory devices.
%\begin{table}[ht!]
%  \setlength{\tabcolsep}{12.0pt}
%  \caption{Nearest neighbor magnetic exchange interaction parameters calculated for $A_4$Co$_3$O$_{9}$ ($A$: La, Nd) using $U_{\rm{eff}}$ = 5.0 eV. As seen, the magnetic exchange parameters calculated for the polar $Cmc2_1$ phase of Nd$_4$Co$_3$O$_{9}$ are almost identical to those calculated for the experimentally observed $Pmcn$ phase of La$_4$Co$_3$O$_{9}$.}
%  \label{tab-exchange}
%  \centering
%  \begin{tabular}{|c|c|c|c|c|c|c|}\hline \hline
%      Layered oxide& Phase & $J_1$ (meV)&$J_2$ (meV)&$J_3$ (meV)&$J_4$ (meV)&$J_5$ (meV) \\
%       \hline  \rule{0pt}{1.2\normalbaselineskip}     La$_4$Co$_3$O$_9$&$Pmcn$&-29.67&-2.79&-3.71&-0.22&-0.86\\   
%       \rule{0pt}{1.5\normalbaselineskip} Nd$_4$Co$_3$O$_9$&$Cmc2_1$&-29.01&-3.78&-3.62&-0.27&-0.96\\   \hline 
%       \hline
%      \end{tabular}
%\end{table}

\begin{table}[ht!]
  \setlength{\tabcolsep}{12.0pt}
  \caption{Nearest neighbor magnetic exchange interaction parameters calculated for $A_4$Co$_3$O$_{9}$ ($A$: La, Nd) using $U_{\rm{eff}}$ = 5.0 eV. As seen, the magnetic exchange parameters calculated for the polar $Cmc2_1$ phase of Nd$_4$Co$_3$O$_{9}$ are almost identical to those calculated for the experimentally observed $Pmcn$ phase of La$_4$Co$_3$O$_{9}$. Since a stronger $J_4$ gives rise to G-AFM ordering, rather than G$^*$-AFM ordering, this further supports the small energy differences between G- vs G$^*$-AFM configurations as being significant. }
  \label{tab-exchange}
  \centering
  \begin{tabular}{|c|c|c|c|c|c|c|}\hline \hline
      Layered oxide& Phase & $J_1$ (meV)&$J_2$ (meV)&$J_3$ (meV)&$J_4$ (meV)&$J_5$ (meV) \\
       \hline  \rule{0pt}{1.2\normalbaselineskip}     La$_4$Co$_3$O$_9$&$Pmcn$&-7.39&-1.50&-1.80&-0.44&-0.43\\   
       \rule{0pt}{1.5\normalbaselineskip} Nd$_4$Co$_3$O$_9$&$Cmc2_1$&-7.22&-2.01&-1.75&-0.51&-0.48\\   \hline 
       \hline
      \end{tabular}
\end{table}
\clearpage
\newpage
\section{S3. L\lowercase{a}$A^\prime_3$F\lowercase{e}$_3$O$_9$ ($A^\prime$: S\lowercase{r}, C\lowercase{a}) layered oxides} \label{Fe}

\subsection{A. CATION ORDERING MODELS}\label{cation-models-Fe}
The high symmetry $Cmcm$ structure contains two inequivalent $A$ sites (Wyckoff position $8g$), each with a multiplicity of two. Therefore, in order to model the La$A^\prime_{3}$Fe$_3$O$_{9}$ layered oxides, we have considered three different cation ordering models as described below. 
\begin{figure}[ht!]
\includegraphics[scale=0.8]{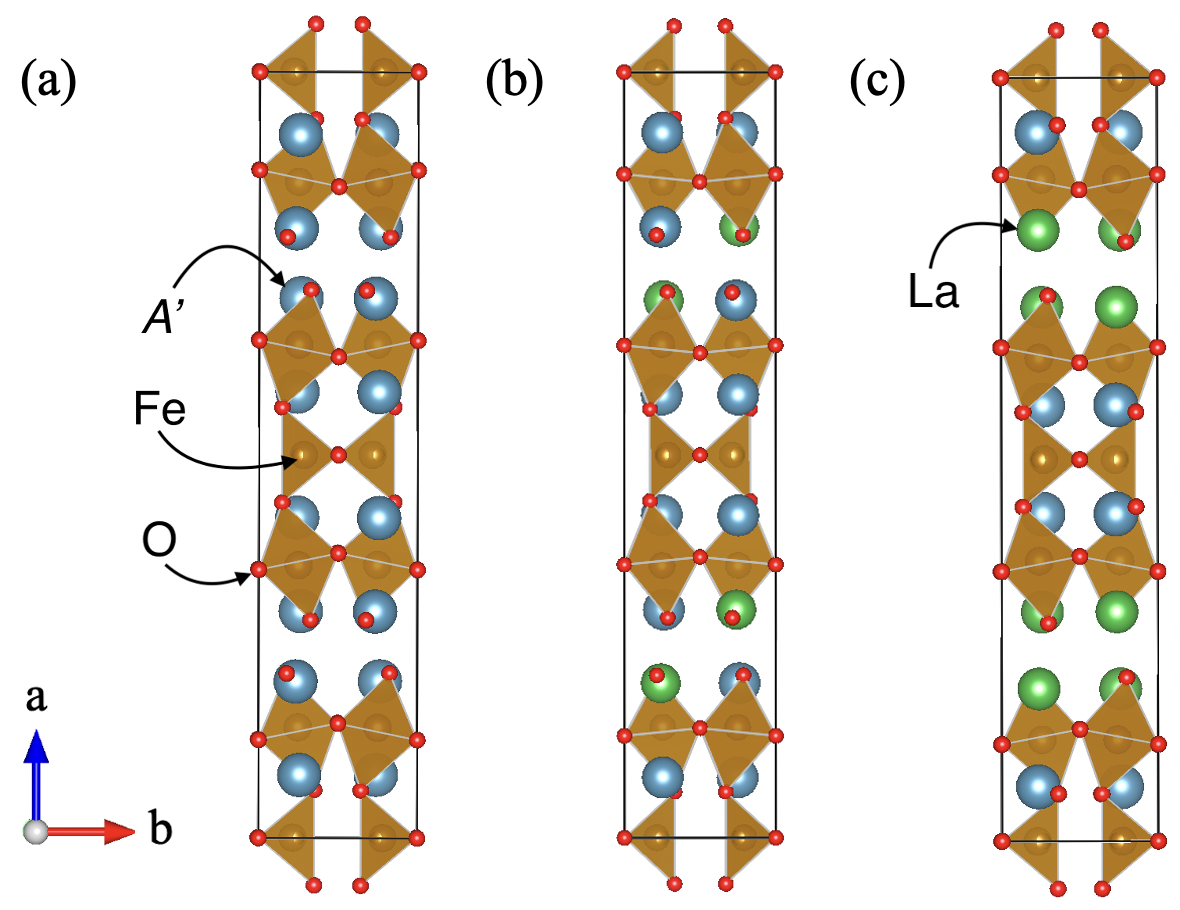}
    \caption{Crystal structures within the (a) electron doped, (b) explicitly doped, and (c) hole doped cation ordering models considered to simulate the La$A^\prime_{3}$Fe$_3$O$_{9}$ (${A^\prime}$ = Sr, Ca) layered oxides.}
\label{cation-model}
\end{figure}

\noindent
\textbf{\textit{Electron doped model:}}
In this model, we considered full occupancy of the ${A^\prime}^{2+}$ ions on the $A$ sites. Four electrons are added to maintain the charge neutrality and $3+$ valency of the Fe ions. The corresponding crystal structure is shown in Fig.~\ref{cation-model}(a).  \\

\noindent
\textbf{\textit{Explicitly doped model:}}
Four La atoms are added explicitly in the charge neutral four formula unit cell. The explicitly doped model breaks the symmetry of the original cell as described in Table~\ref{tab-lattice-Fe}. The lowest energy crystal structure within this model is shown in Fig.~\ref{cation-model}(b). \\

\noindent
\textbf{\textit{Hole doped model:}}
In the hole doped model, La$^{3+}$ and ${A^\prime}^{2+}$ ions fully occupy the $A1$ and $A2$ sites. Our DFT calculations reveal that La prefers to occupy the $A$ sites in the rocksalt layers as shown in Fig.~\ref{cation-model}(c). Similar to the electron doped model, four holes are added to maintain the charge neutrality and $3+$ valency of the Fe ions. 

\clearpage
\subsection{B. STRUCTURAL DETAILS}
\begin{table}[h]
  \setlength{\tabcolsep}{1.5pt}
  \caption{Fully optimized lattice parameters and magnetic moments of the different structural phases of the La$A^\prime_3$Fe$_3$O$_9$ ($A^\prime$: Sr, Ca) oxides calculated with ground state G-AFM magnetic ordering using $U_{\rm{eff}}$ = 4.0 eV for the $3d$ states of Fe. Available experimental values are given in the parentheses for comparison. Here, $\Delta E$ denotes the relative energy of different structural variants. The zero of energy corresponds to the polar $Cmc2_1$ structure. Note that the explicitly doped model breaks the symmetry of the original cell. The reduced symmetry space groups are given in the parentheses. As seen,  the G.S. crystal structures of La$A^\prime_3$Fe$_3$O$_9$ ($A^\prime$: Sr, Ca) compounds are very sensitive to the cation ordering model considered. }
  \label{tab-lattice-Fe}
  \centering
  \begin{tabular}{|c|c|c|c|c c c c|c c|}
    \hline
    \hline\rule{0pt}{1.0\normalbaselineskip}
    Layered&Cation ordering&Phase&$\Delta E$&\multicolumn{4}{c|}{DFT-optimized lattice parameters}& \multicolumn{2}{c|}{Magnetic moments ($\mu_{\rm B}$)} \\\cline{5-8} \cline{9-10} \rule{0pt}{1.0\normalbaselineskip}
     oxide&model& &(meV/f.u.)& $a$ (\AA) & $b$ (\AA)& $c$ (\AA)& $V_{\rm{cell}}$ (\AA$^3$) & $\mu_{\rm{tetra}}$  & $\mu_{\rm{octa}}$  \\
     \hline \rule{0pt}{1.0\normalbaselineskip}
    &Electron&$Cmcm$&437.77&28.7242 &5.6731 &5.6043 &913.25 &3.915 &4.078 \\
     \rule{0pt}{1.0\normalbaselineskip}

       &doped&&& (28.7559$^\dagger$)& (5.5279$^\dagger$)& (5.4582$^\dagger$)&(867.64$^\dagger$)&(3.58$^\dagger$)& (3.99$^\dagger$)\\
     \rule{0pt}{1.0\normalbaselineskip}
    &&$Cmc2_1$&0.00&29.2117&5.6347&5.5304&910.30&3.997&4.064\\
     \rule{0pt}{1.0\normalbaselineskip}
    &&$Pmcn$&$-0.003$&29.2126&5.6346&5.5306&910.34&3.997&4.064\\
    \cline{2-10} \rule{0pt}{1.0\normalbaselineskip}

    &Explicitly&$Cmcm$ ($Pbnm$)&277.62&27.5944&5.6165&5.5314&857.27&3.888&4.094\\
     \rule{0pt}{1.0\normalbaselineskip}
    LaSr$_{3}$Fe$_3$O$_{9}$&doped&$Cmc2_1$ ($Pbn2_1$)&0.00&28.0063&5.5958&5.4822&859.17&3.983&4.085\\
     \rule{0pt}{1.0\normalbaselineskip}
    &&$Pmcn$ ($P2_12_12_1$)&-2.95&27.9938&5.5943&5.4885&859.54&3.983&4.086\\
   \cline{2-10} \rule{0pt}{1.0\normalbaselineskip}
       &Hole&$Cmcm$&97.75&27.0699&5.5038&5.4232&807.99&3.859&4.085\\
     \rule{0pt}{1.0\normalbaselineskip}
    &doped&$Cmc2_1$&0.00&27.4933&5.4816&5.3903&812.36&3.961&4.076\\
     \rule{0pt}{1.0\normalbaselineskip}
    &&$Pmcn$&0.08&27.4959&5.4811&5.3904&812.36&3.961&4.076\\
    \hline \rule{0pt}{1.0\normalbaselineskip}
    &Electron&$Cmcm$&745.56&27.0879&5.6116&5.5485&843.41&3.933&4.093\\
     \rule{0pt}{1.0\normalbaselineskip}
    &doped&$Cmc2_1$&0.00&27.0320&5.5743&5.5447&835.49&4.007&4.090\\
     \rule{0pt}{1.0\normalbaselineskip}
    &&$Pmcn$&9.09&27.0918&5.5727&5.5461&837.31&4.007&4.090\\\cline{2-10}\rule{0pt}{1.0\normalbaselineskip}

    &Explicitly&$Cmcm$ ($Pbnm$)&617.22&26.6787&5.5453&5.4670&808.80&3.913&4.095\\
     \rule{0pt}{1.0\normalbaselineskip}
    LaCa$_{3}$Fe$_3$O$_{9}$&doped&$Cmc2_1$ ($Pbn2_1$)&0.00&26.7691&5.5545&5.4089&804.25&4.010&4.089\\
     \rule{0pt}{1.0\normalbaselineskip}
    &&$Pmcn$ ($P2_12_12_1$)&-0.88&26.7656&5.5554&5.4084&804.19&4.010&4.088\\ 
    \cline{2-10}\rule{0pt}{1.0\normalbaselineskip}
    &Hole&$Cmcm$&524.41&26.6036&5.4483&5.3688&778.18&3.888&4.079\\
     \rule{0pt}{1.0\normalbaselineskip}
    &doped&$Cmc2_1$&0.00&26.7898&5.4570&5.3183&777.49&4.006&4.066\\
     \rule{0pt}{1.0\normalbaselineskip}
    &&$Pmcn$&0.05&26.7909&5.45686&5.3184&777.51&4.006&4.066\\    
    \hline 
    \hline 
    \multicolumn{10}{l}{ $\dagger$ denotes the experimental values measured in Ref.~\cite{LaSr2019}.} \\
\end{tabular}
\end{table}

\clearpage
\subsection{C. ELECTRONIC PROPERTIES}
\begin{figure}[ht!]
\includegraphics[scale=1.6]{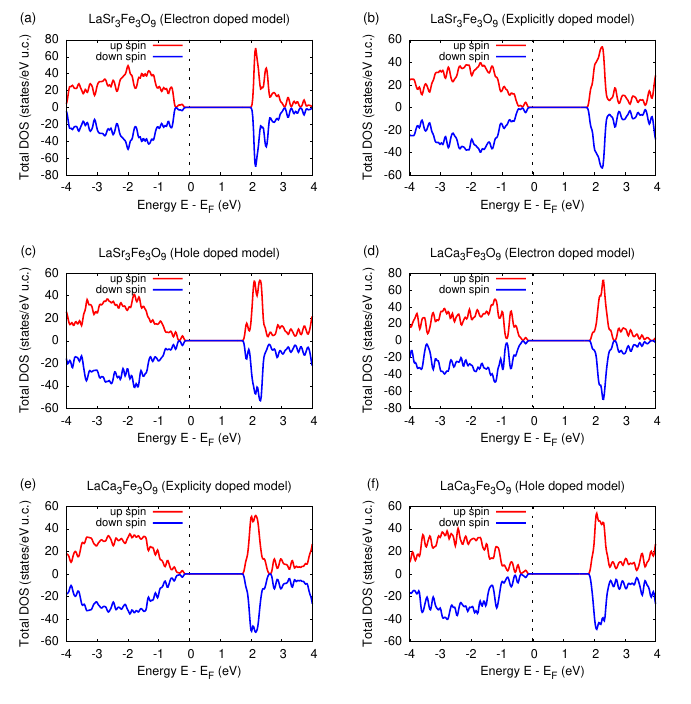}
    \caption{Total density of states (DOS) calculated for the lowest energy phases of La$A^\prime_{3}$Fe$_3$O$_{9}$ ($A^\prime$: Sr, Ca) compounds within different cation ordering models (see Table~\ref{tab-lattice-Fe}) showing the insulating nature of the considered layered oxides. G-AFM magnetic order is considered with $U_{\rm{eff}}$ = 4.0 eV for the $3d$ states of Fe.}
\label{dos-Fe}
\end{figure}

\clearpage
\subsection{D. GROUND STATE MAGNETIC CONFIGURATIONS}
\begin{table}[h]
  \setlength{\tabcolsep}{2.0pt}
  \caption{G.S. magnetic configuration of the La$A^\prime_3$Fe$_3$O$_{9}$ ($A^\prime$: Sr, Ca)  layered oxides calculated for the lowest energy phases given in Table~\ref{tab-lattice-Fe}. We use $U_{\rm{eff}}$ = 4.0 eV for the $3d$ states of Fe. Note that the explicitly doped model breaks the symmetry of the original cell. The reduced symmetry space groups are given in the parentheses. As seen, the G.S. magnetic structures of La$A^\prime_3$Fe$_3$O$_{9}$ ($A^\prime$: Sr, Ca) are very sensitive to the cation ordering model considered.}
  \label{tab-mag-Fe}
  \centering
  \begin{tabular}{|c|c|c|c c c c c c|c|c|c|c|}
    \hline
    \hline\rule{0pt}{1.2\normalbaselineskip}
    Layered&Model&Phase&\multicolumn{6}{c|}{Relative energy in meV/f.u. }&G.S.&Polar&Magnetic &wFM\\\cline{4-9}  \rule{0pt}{1.2\normalbaselineskip}
     oxide& && FM & A-AFM & C-AFM&C$^*$-AFM& G-AFM&G$^*$-AFM&symmetry&mode&modes& \\
     \hline \rule{0pt}{1.5\normalbaselineskip}       &Electron &$Pmcn$&606.24 &519.45 &50.59 &50.62&0.03 &0.00&$Pm^\prime cn^\prime$&\xmark&$m\Gamma^+_4(a)$&\cmark \\
    \rule{0pt}{1.5\normalbaselineskip} 
    &doped&& && & &&&&&$mY^-_3(a)$& \\
    \cline{2-13} \rule{0pt}{1.5\normalbaselineskip}  
    LaSr$_{3}$Fe$_3$O$_{9}$&Explicitly&$Pmcn$ &735.62 &567.99&118.78&118.66&0.00 &0.09&$Pm^\prime cn$&\xmark&$m\Gamma^-_3(a)$&\xmark \\
    \rule{0pt}{1.5\normalbaselineskip} 
    &doped&$(P2_12_12_1)$& & && &&&$(P2_12^\prime_12^\prime_1)$&&$mY^+_4(a)$& \\ 
    \cline{2-13} \rule{0pt}{1.5\normalbaselineskip}
    &Hole&$Cmc2_1$&898.54 &646.10 &190.27&189.70&0.00 &0.50&$Cm^\prime c2_1^\prime$&\cmark&$m\Gamma^+_4(a)$&\cmark \\
    \rule{0pt}{1.5\normalbaselineskip} 
    &doped&& & && &&&&&$m\Gamma^-_3(a)$& \\ 
    \hline \rule{0pt}{1.5\normalbaselineskip} 
    &Electron &$Cmc2_1$&571.17 &448.58 &91.99 &92.06&0.50&0.00&$P_Cbc2_1$&\cmark&$mY^+_4(a)$&\xmark \\
     \rule{0pt}{1.5\normalbaselineskip} 
    &doped&& & && &&&&&$mY^-_3(a)$& \\
    \cline{2-13} \rule{0pt}{1.5\normalbaselineskip}  
    LaCa$_{3}$Fe$_3$O$_{9}$&Explicitly&$Pmcn$ &753.18 &542.00&168.29&168.01&0.00 &0.24&$Pm^\prime cn$&\xmark&$m\Gamma^-_3(a)$&\xmark \\
    \rule{0pt}{1.5\normalbaselineskip} 
    &doped&$(P2_12_12_1)$& & & &&&&$(P2_12^\prime_12^\prime_1)$&&$mY^+_4(a)$& \\ 
    \cline{2-13} \rule{0pt}{1.5\normalbaselineskip}
    &Hole&$Cmc2_1$&918.49 &609.81 &255.93&255.25&0.00 &0.60&$Cm^\prime c2_1^\prime$&\cmark&$m\Gamma^+_4(a)$&\cmark\\
    \rule{0pt}{1.5\normalbaselineskip} 
    &doped&& && & &&&&&$m\Gamma^-_3(a)$& \\  
    \hline 
    \hline
\end{tabular}
\end{table}

\begin{table}[h]
  \setlength{\tabcolsep}{6.0pt}
  \caption{Relative energies of the magnetic easy axis directions calculated for the lowest energy phases of La$A^\prime_3$Fe$_3$O$_{9}$ ($A^\prime$: Sr, Ca) within the electron doped cation model. Ground state G-AFM spin ordering is considered for the noncollinear calculations including SOC. $U_{\rm{eff}}$ = 4.0 eV is used for the $3d$ states of Fe.}
  \label{tab-easydir-Fe}
  \centering
  \begin{tabular}{|c|c|c c c|}
    \hline
    \hline\rule{0pt}{1.2\normalbaselineskip}
    Layered&Phase&\multicolumn{3}{c|}{Relative energy in meV/f.u. }\\\cline{3-5}  \rule{0pt}{1.2\normalbaselineskip}
     oxide& & along $a$ &along $b$ & along $c$  \\
     \hline \rule{0pt}{1.2\normalbaselineskip}   
    LaSr$_{3}$Fe$_3$O$_{9}$&$Pmcn$& 0.42&0.04 &0.00 \\
    \rule{0pt}{1.5\normalbaselineskip}     
    LaCa$_{3}$Fe$_3$O$_{9}$&$Cmc2_1$& 0.32& 0.02&0.00  \\
    \hline 
    \hline
\end{tabular}
\end{table}

\clearpage
\subsection{E. PHONON SPECTRA AND METASTABLE PHASES}

\begin{figure}[ht!]
\includegraphics[scale=0.65]{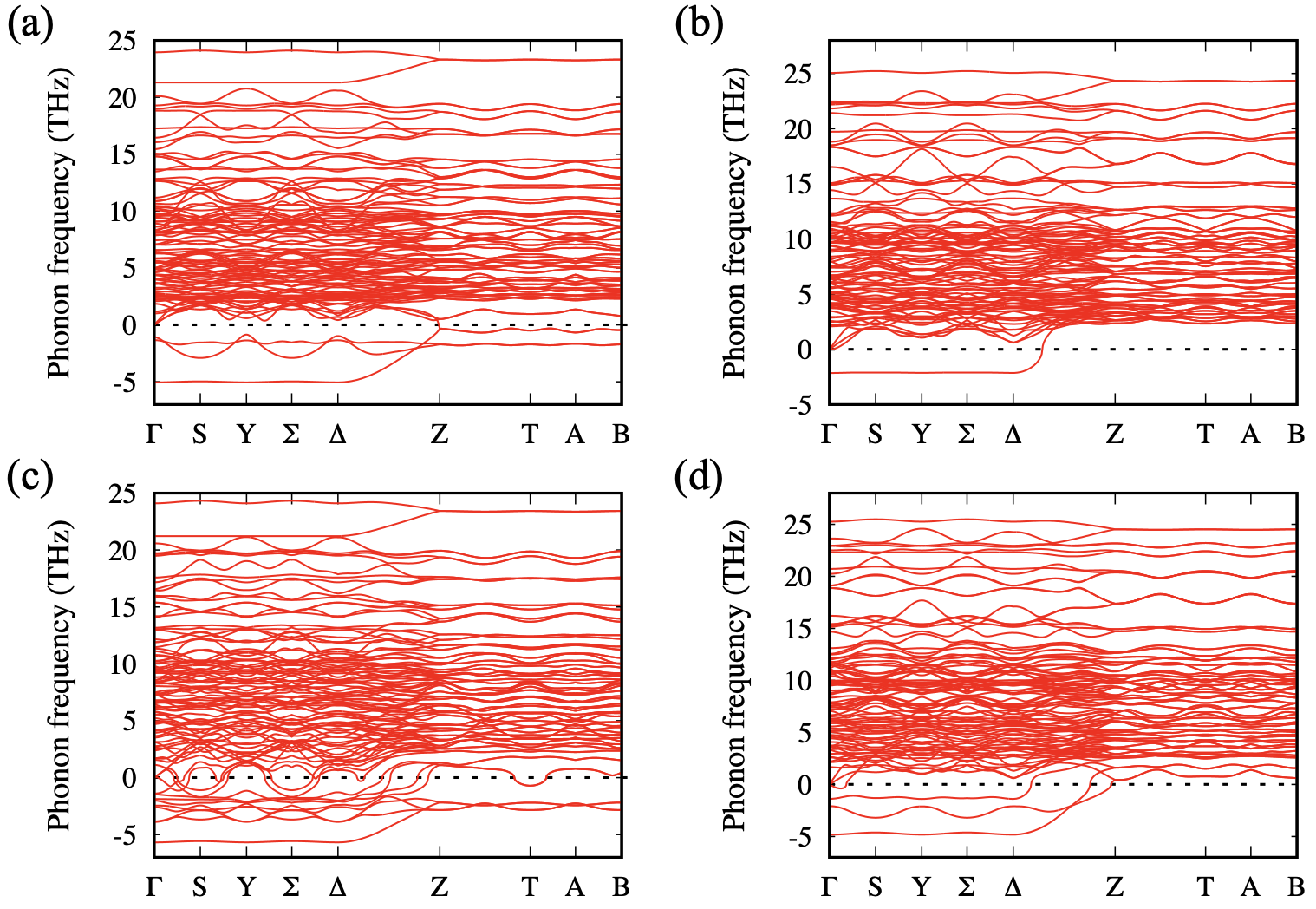}
    \caption{(a)$-$(b) Phonon dispersion curves along the $\Gamma (0, 0, 0)-S(0, 0.5, 0)-\Sigma (0.25, 0.25, 0)-Y(0.5, 0.5, 0)-\Delta(0.75, 0.25, 0) -Z(0, 0, 0.5)-T(0.5, 0.5, 0.5)-A(0.25, 0.25, 0.5)-B(0.75, 0.25, 0.5)$ direction of the BZ (in the primitive basis) calculated with G-AFM magnetic order for the high symmetry $Cmcm$ phase of LaSr$_{3}$Fe$_3$O$_{9}$ in the electron doped (left) and hole doped (right) models. (c)$-$(d) Corresponding phonon bands of LaCa$_{3}$Fe$_3$O$_{9}$ in the electron doped (left) and hole doped (right) models. $U_{\rm{eff}}$ = 4.0 eV is used for the $3d$ states of Fe. Both the models show qualitatively similar features in the phonon spectra with a strongly unstable flat phonon branch along the $\Gamma-S-\Sigma-Y-\Delta$ direction in the BZ associated with the tetrahedral chain ordering distortions. The phonon spectra within explicitly doped model are not shown as the symmetry of the simulated unit cell differs from the $Cmcm$ symmetry of the original parent cell.}
\label{phonon-Fe}
\end{figure}

\clearpage
\begin{table}[h]
  \setlength{\tabcolsep}{4.0pt}
  \caption{Relative energies of the different metastable phases of La$A^\prime_{3}$Fe$_3$O$_{9}$ ($A^\prime$: Sr, Ca) within the electron doped model arising from the relative ordering of the tetrahedral twisting patterns shown in Fig.~\ref{superstruct}. We fully relaxed the structures with G-AFM spin order using $U_{\rm{eff}}$ = 4.0 eV for the $3d$ orbitals of Fe. Note that the polar $Cmc2_1$ phase is the ground state of LaCa$_{3}$Fe$_3$O$_{9}$ which allows for the $\Gamma$-point switching scheme described in Fig.~1(b) of the main text.}
  \label{tab-sup-Fe}
  \centering
  \begin{tabular}{|c|c|c| c |c |}
    \hline
    \hline\rule{0pt}{1.2\normalbaselineskip}
    Phase&Irrep of primary &Irreps of secondary order&\multicolumn{2}{c|}{Relative energy in meV/f.u. }\\\cline{4-5}  \rule{0pt}{1.5\normalbaselineskip}
     &order parameter&order parameters&LaSr$_{3}$Fe$_3$O$_{9}$&LaCa$_{4}$Fe$_3$O$_{9}$  \\
     \hline \rule{0pt}{1.5\normalbaselineskip}   
    %&$Cmcm$&524.82&179.11 &20.89 &0.00 &0.09&$Pm^\prime cn^\prime$&\xmark&\cmark\\
     %\rule{0pt}{1.5\normalbaselineskip}
    $Cmc2_1$&$\Gamma^-_2(a)$&$\Gamma^+_1(a)$&3.52&0.00 \\
     \hline \rule{0pt}{1.5\normalbaselineskip} 
    $Pmcn$&$Y^-_2(a)$&$\Gamma^+_1(a)$&3.51&9.09 \\
    \hline \rule{0pt}{1.5\normalbaselineskip}
    $Pmab$&$\Sigma_3(0, a)$&$\Gamma^+_1(a)$&227.60&177.26 \\
      \rule{0pt}{1.5\normalbaselineskip}
     &&$Y^+_1(a)$&& \\
     \hline \rule{0pt}{1.5\normalbaselineskip}
     $Pbca$&$\Sigma_3(a, -a)$&$\Gamma^+_1(a)$&8.22&19.23 \\
     \rule{0pt}{1.5\normalbaselineskip}
     &&$Y^-_3(a)$&& \\
     \hline \rule{0pt}{1.5\normalbaselineskip}
     $Pmcb$&$\Delta_3(a, 0)$&$\Gamma^+_1(a)$&226.22&387.03\\
      \rule{0pt}{1.5\normalbaselineskip}
     &&$Y^+_1(a)$&& \\
     \hline \rule{0pt}{1.5\normalbaselineskip}
     $Pmnb$&$\Delta_3(a, -a)$&$\Gamma^+_1(a)$&0.00&93.95 \\
     \rule{0pt}{1.5\normalbaselineskip}
     &&$Y^-_4(a)$&& \\
     \hline \rule{0pt}{1.5\normalbaselineskip}
    $Cc2m$&$Z_2(a, 0)$&$\Gamma^+_1(a)$&414.31&614.35 \\
      \rule{0pt}{1.5\normalbaselineskip}
     &&$\Gamma^-_4(a)$&& \\
    \hline 
    \hline
\end{tabular}
\end{table}

%\end{document}